\PassOptionsToPackage{sort&compress,numbers,super}{natbib}

\documentclass[prd,superscriptaddress,amsfonts,amssymb,amsmath,showpacs]{revtex4-2}
\usepackage[utf8]{inputenc}
\usepackage{siunitx}
\usepackage{url}
\usepackage{tensor}
\usepackage{longtable}
\usepackage{float}
\usepackage{mathtools}
\usepackage{multirow}
\usepackage{amssymb}
\usepackage{amsmath}
\usepackage{makecell,tabularx,multirow}
\usepackage{color, colortbl}
\usepackage{hyperref}
\usepackage{url}
\usepackage{glossaries}

\usepackage{achemso}

\makeglossaries
\usepackage{adjustbox}
\usepackage{tikz}
\usetikzlibrary{positioning}
\usetikzlibrary{shapes.geometric, arrows}
\usepackage[pages=some]{background}
\usepackage{hyperref}
\usepackage{accents}
\usepackage{todonotes}
\usepackage{eqnarray}
\usepackage{makecell}

\usepackage{makecell}
\usepackage[margin=1in]{geometry}
\hypersetup{
    colorlinks=true,
    linkcolor=black,
    filecolor=magenta,
    citecolor=red
}
\usepackage{comment}
\usepackage{booktabs}

\newcolumntype{C}{>{\centering\arraybackslash}X}
\newcolumntype{x}[1]{>{\centering\arraybackslash\hspace{0pt}}p{#1}}

\usepackage[margin=1in]{geometry}

\tikzstyle{startstop} = [rectangle, rounded corners, 
text width=7em, minimum height=1cm,text centered, draw=black, fill=red!30]
\tikzstyle{io} = [trapezium, trapezium left angle=70, trapezium right angle=110, 
text width=7em, minimum height=1cm, text centered, draw=black, fill=blue!30]
\tikzstyle{process} = [rectangle, 
text width=7em, minimum height=1cm, text centered, draw=black, fill=orange!30]
\tikzstyle{decision} = [diamond, minimum height=1cm, text centered, text width=5.5em, node distance=3cm, draw=black, fill=green!30]
\tikzstyle{arrow} = [thick,->,>=stealth]

\makeglossaries

\begin{document}

\title{Testing Planck 2020 and DESI data on $w$CDM Models}

\author{Kathleen Sammut}
\email{kathleen.sammut.19@um.edu.mt, ksamm09@um.edu.mt, kathsammut27@gmail.com}
\affiliation{Institute of Space Sciences and Astronomy, University of Malta, Msida, Malta}
\affiliation{Department of Physics, University of Malta, Msida, Malta}

\begin{abstract}
This work delves into the dynamical dark energy models of the $w$CDM parameterisation that are defined by their equation of state by comparing different, well-known parameterisation models, in an attempt to lessen the tensions of $H_0$ and $\sigma_8$ by using the latest observational data. This research also tested the newer Planck likelihood and comparing it to the previously released dataset of Planck 2018 and the newer \gls{BAO} data of \gls{DESI}. The data that was used were: the Cosmic Microwave Background (\gls{CMB}) data of Planck 2018 and Planck 2020 data; Cosmic Chronometers (\gls{CC}), a sample of Supernovae Type Ia; and Baryonic Acoustic Oscillations (\gls{BAO}). A Bayesian analysis was performed to produce the results needed for the analysis. From the analyses, the best-fit values of the parameters show that almost all models are in favour of a phantom Universe when early-time data was used and favours a quintessence Universe when combinations of early-time and late-time data was considered. This study also showed that the new tested data constrained the models better than the previous ones, while also showing that the observation data supports the $\Lambda$CDM model over the dynamical dark energy models.\\

\end{abstract}

\maketitle

\renewcommand{\thesection}{\arabic{section}}
\renewcommand{\thesubsection}{\arabic{section}.\arabic{subsection}}
\renewcommand{\thesubsubsection}{\arabic{section}.\arabic{subsection}.\arabic{subsubsection}}


\section{Introduction}\label{sec:intro}

During this last decade, the cosmic tension has been continuously debated in modern cosmology as discrepancies between the values of the Hubble parameter, $H_0$, \cite{H0Tensions} and the amplitude of the matter power spectrum, $\sigma_8$, \cite{Planck2018} derived either indirectly, through the use of models or directly, through standard candles \cite{Supernova,Supernova2}, still persist. This gave rise to the possibility of an alternative model to the standard model. According to observational evidence, the $\Lambda$ Cold Dark Matter ($\Lambda$CDM) \cite{LCDM} cosmology is considered the most well-suited model for explaining the Universe. It presumes that the cosmological constant, $\Lambda$, is responsible for driving the accelerated expansion of the Universe. However, despite its success in describing the late accelerating phase, $\Lambda$CDM faces several significant challenges. Apart from the tensions with the $H_0$ \cite{H0Tensions2,H0Tensions3} and $\sigma_8$, the standard model also has problems regarding the cosmological constant problem, which highlights the discrepancy between the observed small value of the cosmological constant and the large theoretical value predicted by quantum field theory. Furthermore, $\Lambda$ cosmology fails to address the cosmic coincidence problem, further emphasising the need for alternative explanations or modifications to the model. This further pushes the scientific community to search for an alternative model that might lessen or even remove the problems found in the standard model of cosmology \cite{ModifiedGravity3}.\\

To explain the phenomenon of $H_0$, two main approaches have been proposed: one involves keeping to the theory of general relativity while introducing new components that go beyond the Standard Model of Particle Physics \cite{ModifiedGravity,ModifiedGravity2}; the other involves constructing modified theories of gravity that contain additional degrees of freedom from which one can derive the Universe's expansion rate. Since the Universe is assumed to be a fluid, both approaches lead to a value of the expansion rate of the Universe that is affected by the evolution of the dark energy equation of state, $w$. This parameter $w$, represents the ratio of the dark energy's pressure, $p$, to its energy density, $\rho$. The standard model takes the assumption that $w=-1$, meaning that the Universe behaves like a dark energy fluid due to it being negative. The simplest generalisation of the $\Lambda$CDM model is to relax the assumption that $w$ is $-1$ throughout the history of the Universe, known as the $w$CDM model \cite{ModelsEx,4parameters2}. The first approach to this is taking $w$ to be a constant, but $w \neq 1$ \cite{ModelEx3}. However, a better generalisation of the standard model would be one that considers the equation of state to be time-dependent. This can be done by parameterising $w$ in terms of the cosmological redshift with the use of different functional forms. This gives more freedom to reconstruct the expansion history of the Universe with respect to the observational data.\\

In the literature, many different dark energy models can be found that fit the observational data: there are one-parameter models, such as the constant model; two-parameter models, namely, Chevallier-Polarski-Linder parametrization \cite{CPL1,CPL2}, Jassal-Bagla-Padmanabhan parametrisation \cite{JBP, myModelsPaper}, Logarithmic parametrisation \cite{GE}, and oscillatory parameterisation models \cite{Oscillmodel, myOscillpaper}; three-parameter models \cite{3-4Parameters}; and four-parameter parameterizations \cite{3-4Parameters,4parameters,4parameters2}. $w$CDM models received significant attention when recent studies suggested that observational data favoured a dynamic dark energy equation of state. As precision measurements keep advancing and observational data becomes more precise, the need to explore different dark energy equation of state models becomes more pressing. \\

New cosmic microwave background data (\gls{CMB}) and Baryonic Acoustic Oscillations (\gls{BAO}) data sets, Planck 2020 and \gls{DESI}, were released in 2022 and 2024, respectively. Since these two data sets are still considered new, rigorous testing needs to be done on them to make sure that these data sets can be trusted for the testing of models. In this study, the examination of six popular dark energy models that are characterised by their equation of state will be carried out with the help of \gls{CMB} data of Planck 2018 \cite{Planck2018,Planck2018_2} and Planck 2020 \cite{Planck2020}, late-time data of \gls{CC} \cite{CC} and \gls{SN$+$}\gls{SH0ES} \cite{Pantheon+,SH0ES} and \gls{BAO} \cite{BAO}
data. Beyond deriving cosmological constraints on these proposed parametrisations, the aim is to evaluate whether these models can address the Hubble constant and the sigma tensions. This study will also test the new Planck and DESI data by comparing them to their predecessors.\\

This paper has been sectioned in the following way. In section \ref{sec:wCDM}, the $w$CDM model will be explained by its background and perturbation equations as well as showcasing the chosen models by their equation of state. In section \ref{sec:data}, the chosen observation data that was used in the analysis will be presented. Section \ref{sec:results}, displays all the results that were obtained from this study. In section \ref{sec:analysis}, the analysis of the observational constraints on the parameter space of the $w$CDM models and the comparison of the data set will be carried out. Finally, section \ref{sec:conclusion}, closes off with a brief summary of all the results and conclusions that were made through this study. \\

\section{$w$CDM Models} \label{sec:wCDM}

The background and perturbation equations are the integral part of any model. The background equations assume an isotropic and homogeneous Universe. However, at lower scales or high modes, the cosmological principle no longer applies; therefore, the perturbation equations handle the fluctuations that are found in the Universe, such as density and gravitation. Starting from the background equations, the metric is what is known as the Friedmann-Lema$\hat{i}$tre-Robertson-Walker (\gls{FLRW}) metric, 
\begin{equation}
    ds^2 = -dt^2 +a^2(t)\left[\frac{dr^2}{1-kr^2} + r^2(d\theta^2 + \sin^2\theta d\phi^2) \right]~,
\end{equation}
where $a(t)$ represents the scale factor and $t, \theta, \phi$ are to comoving coordinates in polar form. $k$ can be $-1,0$ and $+1$ depending on whether an open, flat or closed Universe is considered. In this study, a flat Universe is assumed a priori as it is favoured by observational data. By using the FLRW metric, the Einstein Field equations, and the stress-energy momentum tensor, the first and second Friedmann equations can be derived as,
\begin{equation}
    H^2 = \frac{8\pi G}{3}\rho_{tot}
\end{equation}
where $H$ is the Hubble parameter and is defined as $H = \frac{\dot{a}}{a}$, where the dot represents the derivative with respect to cosmic time $t$. $\rho_{tot}$ represents the total density of the Universe for all the different types of matter. The second Friedmann equation is derived from the Einstein field equations and the stress-energy tensor and has the form,
\begin{equation}
    \frac{\ddot{a}}{a} = \frac{-4\pi G}{3}(\rho_i +3P_{tot})~,
\end{equation}
where $P_{tot}$ is the total pressure for all different types of matter. Each type of matter is assumed not to mutually interact, then each type of matter is separately conserved. Therefore, each type of matter satisfies the continuity equation, which is the conservation of energy for an expanding Universe, and is derived from the first law of thermodynamics as,
\begin{equation}
    \dot{\rho} + 3H(1+w_i)\rho_i = 0~,
\end{equation}
where $w_i \equiv \frac{p_i}{\rho_i}$ is the equation of state parameter and $i$ denotes the type of matter. From the continuity equation, the density of radiation, baryonic matter, cold dark matter, and dark energy can be found. Since radiation has $w_r= \frac{1}{3}$, then $\rho_r \propto \left( \frac{a}{a_0}\right)^{-4}$. In the case of baryonic and cold dark matter, $w_b =0$ and $w_c=0$, therefore, $\rho_b\propto \left(\frac{a}{a_0}\right)^{-3}=\rho_c$ \cite{myModelsPaper} where $a_0$ is the value of the scale factor at current times. Since, dark energy is no longer taken as a constant, like in the standard model, the density of dark energy also needs to be calculated by using the continuity equation \cite{myOscillpaper}.
\begin{equation}
    \rho_{\Lambda}=\rho_{\Lambda_0}\left(\frac{a}{a_0}\right)^{-3}\exp\left(-3\int^{a}_{a_0}\frac{w_{\Lambda}(a')}{a'}da'\right)
\end{equation}
By substituting the densities into the Friedmann equation, we obtain the complete Friedmann equation of $w$CDM models,
\begin{equation}
    H^2 = \frac{8\pi G}{3}\left[\rho_{r_0}a^{-4}+\rho_{b_{0}}a^{-3} + \rho_{c_{0}}a^{-3} + \rho_{\Lambda_0}\left(\frac{a}{a_0}\right)^{-3}\exp\left(-3\int^{a}_{a_0}\frac{w_{\Lambda}(a')}{a'}da'\right)\right]~,
\end{equation}
where $\rho_{r_0}$, $\rho_{b_0}$, $\rho_{c_0}$ and $\rho_{\Lambda_0}$ are the present day densities of radiation, baryonic matter, cold dark matter and dark energy, respectively. Moving on to the perturbation equations, even though the $w$CDM model has the same form of perturbation equations as the standard model \cite{PerturbationTheory}. However, the $w$CDM model still impacts the perturbed space. The perturbed \gls{FLRW} metric when taking the synchronous gauge is represented as,
\begin{equation}
        ds^2 = a^2(\tau)[-d\tau ^2 + (\delta_{ij} +h_{ij})dx^i dx^j]~,
\end{equation}
where $\tau$ is the conformal time, $h_{ij}$ and $\delta_{ij}$ are respectively the perturbed and unperturbed metric \cite{myOscillpaper}, which can include scalar, vector, and tensor components. From the perturbed \gls{FLRW} and the stress-energy momentum tensor, the perturbed continuity equation and the perturbed Euler equation can be derived \cite{Perturbationtheory2,PerturbationTheory3},
\begin{eqnarray}
   {\delta}'_i&=&-(1+w_i)\left(\theta_i+\frac{{h}'}{2}\right) - 3\mathcal{H}\left(c^2_s-w_i \right)\delta_i -9\mathcal{H}^2 \left(\frac{\delta P_i}{\delta \rho _i}-c^2 _{a,i}\right)(1+w_i)\frac{\theta_i}{k^2}~\label{pertcontin},
    \\
    {\theta'}_i &=& - \mathcal{H}\left(1-3\frac{\delta P_i}{\delta \rho_i}\right)\theta_i + \frac{\delta P_i}{\delta \rho_i}\frac{1}{1-w_i} k^2 \delta_i -k^2 \sigma_i~\label{pertEuler}~.
\end{eqnarray}
In Eq.~\ref{pertcontin} and Eq.~\ref{pertEuler}, the prime denotes differentiation with respect to the conformal time $\tau = \int \frac{dt}{a(t)}$. The density perturbation is represented by $\delta_i = \frac{\delta \rho_i}{\rho_i}$, the conformal Hubble parameter is represented as $\mathcal{H} = \frac{a'}{a}$, $h=h_{ij}$ is the trace of the metric perturbations $h_{ij}$, the divergence of the ith fluid velocity is denoted by $\theta_i \equiv ik^j v_j$, $c^2_s$ is the sound speed for an imperfect fluid and it is defined as $c^2_s \equiv \frac{\delta P_i}{\delta \rho_i}$, $\sigma_{i}$ is the anisotropic stress of the fluid, which will be neglected in the analysis and $c^{2}_{a,i}$ is the adiabatic speed of sound \cite{soundspeed2}. Therefore, $c^2_{a,i}$ can be related to $w_i$ with the equation \cite{soundspeed},
\begin{equation}
    c^2_{a,i} = w_i - \frac{w'_i}{3\mathcal{H}(1+w_i)}~\label{AdiabaticSpeedSound}.
\end{equation}
Thus, for any equation of state $w(a)$, the Friedmann equation and the adiabatic speed of sound \cite{soundspeed3} can be calculated. In this study, four well-known $w$CDM parameterisations were considered;
\begin{itemize}
    \item \textbf{Parameterisation 1:} The simplest generalisation considered in this study involved assuming that $w(a)$, the dark energy equation of state parameter, is a constant value that is not equal to $-1$. This slight deviation from the $\Lambda$CDM model introduces a more flexible framework for understanding dark energy. The equation of $w(a)$ that represents this representation is given as:
    \begin{equation}
        w_{wCDM}~(a)=w_{0,wCDM}~.
    \end{equation}
    This generalisation is often referred to as the constant model and serves as a baseline for comparison with dynamical dark energy equation of state models.
    \item \textbf{Parameterisation 2:} For the exploration of non-linear dynamical dark energy equation of state, a quadratic parameterisation was taken as,
    \begin{equation}
        w_{JBP}~(a)=w_{0,JBP}+w_{a,JBP}~a(1-a)~.
    \end{equation}
    This parameterisation was proposed by Jassal-Bagla-Padmanabhan \cite{JBP} and is known as the \gls{JBP} parameterisation. This model incorporates a quadratic term that accounts for more complex dynamical behaviour, potentially offering a better fit to observational data.
    \item \textbf{Parameterisation 3:} A logarithmic model for the dark energy equation of state was proposed by G. Efstathiou \cite{GE}, describing a different type of evolution for $w(a)$. The equation is expressed as,
    \begin{equation}
        w_{GE}~(a)=w_{0,GE} - w_{a,GE}\ln(a)~.
    \end{equation}
    This parameterisation was used in this study and will be referred to as the logarithmic model. This logarithmic parameterisation diverges from the polynomial approaches, focusing instead on a logarithmic evolution. 
    
    \item \textbf{Parameterisation 4:} Lastly, an alternative parametrisation is an oscillatory model for the dark energy equation of state. Proposed by Zhang and Ma \cite{Oscillmodel}, suggests an oscillatory equation of state and is expressed as,
    \begin{equation}
       w_{OSCILL}(a) = w_{0,{OSCILL}} +w_{a,{OSCILL}} \left[ a \sin\left(\frac{1}{a}\right) -\sin(1) \right]~.
    \end{equation}
    This oscillatory parameterisation explores the possibility of a $w(a)$ that exhibits periodic oscillations over the cosmic time.
\end{itemize}
In all the above parameterisations, the term $w_{0,i}$ represents the current value of $w(a)$ for the specific model $i$, while $w_{a,i}$ characterize the evolution of $w(a)$ as a function of the scale factor. These models share the property of having a constant dark energy equation of state at the early Universe, to then transition to more dynamic behaviour as the Universe evolves and moves to late times. This diversity in parameterisations enables a comprehensive analysis of different possible dark energy behaviours.

\section{Observational Data Samples}  \label{sec:data}

In this section, the cosmological data sets that were used in this study will be outlined, describing the statistical methodology employed to constrain the dynamical dark energy models considered in this study.

\begin{itemize}
    \item \textbf{Cosmic Microwave Background (\gls{CMB}) data from Planck 2020:} The latest Planck \gls{CMB} data was taken in this study, giving rise to the opportunity for testing the newer dataset. \gls{CMB} temperature and polarisation anisotropies are taken along with their cross-correlations, including high and low multiples $l$. The low-$l$ temperature anisotropies (TT) data were taken from the Lillipop data, using the multipole range of $l<30$. Since Lillipop only has temperature data, the polarisation data of Planck 2018 \cite{Planck2018}, for low multipoles, was used for a complete analysis of the data and the chosen models. Hillipop was used for the high multipole range data of $l=30-2500$ \cite{Planck2020}. This includes: polarisation anisotropies; temperature anisotropies; and their cross-correlations (TT$+$ET$+$EE). The data was accessed via the publicly available Planck likelihood, which marginalises over various nuisance parameters associated with the measurements. For more information, we refer the reader to \cite{Planck2020}. Throughout this study, this dataset will be referred to as \gls{PR4}.
    
    \item \textbf{Cosmic Microwave Background (\gls{CMB}) data from Planck 2018:} In addition to the latest \gls{CMB} measurements in temperature and polarisation from Planck 2020, earlier observations from Planck 2018 were incorporated for a comparison of the analysis. This comparison enables us to assess the differences and impacts arising from the old and new \gls{CMB} datasets. The complete data on temperature, polarisation, and cross-correlations of Planck 2018 will be taken for the low and high multipole ranges. Similar to \gls{PR4}, this \gls{CMB} data set takes $l$ to be in the range of $2-29$ for low-$l$ EE and low-$l$ TT. On the other hand, for high-$l$ TT$+$TE$+$EE, Planck 2018 uses the range of $ 30 \leq \textit{l} \leq 2508$ for TT and $ 30 \leq \textit{l} \leq 1996$ for TE and EE. This dataset will be referred to as \gls{PR3} throughout this study.

    \item \textbf{Cosmic Chronometers (\gls{CC}) data:} The latest Cosmic Chronometers data was used, which consists of 30 data points of $H_0$ across the range of $ 0 < z < 2$ \cite{CC_3,CC_4}, utilizing massive, passively evolving galaxies in the Universe, referred to as cosmic chronometers. These measurements are highly precise, obtained through the spectroscopic method, and are independent of any specific model. This data was compiled in \cite{CC_Comp, CC}. The $\chi^2$ statistic for the cosmic chronometer dataset is expressed as follows:
    \begin{equation}
        \chi^2_{CC} = \sum^{30}_{i=1} \frac{(H(z_i)-H_{th}(z_i))^2}{\sigma^2_i},
    \end{equation}
    where each $z_i$ with its corresponding uncertainty $\sigma_8$ can be found in Table 4 of \cite{CC_Comp}. This data will be referred to as the \gls{CC} data.

    \item \textbf{Supernovae Type Ia from Pantheon+ (\gls{SN$+$}) and Supernova H0 for the Equation of State (\gls{SH0ES}):} Pantheon+ (\gls{SN$+$}) \cite{SN+} is an updated data set of Pantheon (\gls{SN}). It is a compilation of 1701 light curves of 1550 Type Ia supernovae data points ranging from $0.001 \leq z \leq 2.26$. The updated \gls{SN$+$} data set incorporates improvements in calibration, reduction of systematic errors, and a more comprehensive range of supernovae compared to its predecessor. A prior from the Supernova H0 for the Equation of State (\gls{SH0ES}) project \cite{SH0ES} was considered. This prior obtained a value of $73.04 \pm 1.04$ Km s$^{-1}$Mpc$^{-1}$ for $H_0$, which was calculated using Cepheid variable stars in the host galaxy of 42 Type Ia supernovae. The incorporation of the \gls{SH0ES} prior to the \gls{SN$+$} data set combines high-redshift supernova data with local measurements, leading to more accurate and reliable measurements of $H_0$. The \gls{SN$+$} data with the \gls{SH0ES} prior has the combined likelihood of,
    \begin{equation}
        \chi^2=\Delta D^T ( Cov^{SN+}_{stat+syst}+Cov^{SH0ES}_{stat+syst})^{-1}\Delta D~,
    \end{equation}
    where $Cov^{SN+}_{stat+syst}$ and $Cov^{SH0ES}_{stat+syst}$ are the statistical and systematic covariance matrix of Pantheon+ and \gls{SH0ES}, respectively. On the other hand, $\Delta D$ is the difference between the distance of \gls{SN$+$} and either the distance of the host galaxies calculated from \gls{SH0ES} or the distance calculated by the predicted model. In this study, the Pantheon+ data will be used with the SH0ES prior and will be referred to as \gls{SN$+$SH0ES}.
    
    \item \textbf{Baryon Acoustic Oscillation (\gls{BAO}):} For the \gls{BAO} data, the ratio $\frac{r_s}{D_V}$ serves as a “standard ruler,” where $r_s$ represents the comoving sound horizon at the baryon drag epoch, and $D_V$ denotes the effective distance derived from $D_A$. The angular diameter distance $D_A$ and the Hubble parameter $H$ are connected via the equation below:
    \begin{equation}
        D_V(z)=\left[(1+z^2)D_A(a)^2\frac{z}{H(z)}\right]^{\frac{1}{3}}~.
    \end{equation}
    This dataset uses data from large-scale surveys such as the Sloan Digital Sky Survey (\gls{SDSS}) \cite{SDSS} of DR10 and DR11 \cite{SDSSDR10,SDSSDR11}, the Dark Energy Survey (\gls{DES}) \cite{DES2016} and the Baryon Oscillation Spectroscopic Survey (\gls{BOSS}) DR10 \cite{BOSS2020,BOSSDR10}. The $\chi^2$ function of the \gls{BAO} dataset is as follows \cite{BAO},
    \begin{equation}
        \chi^2_{BAO}=\sum_i \frac{\left[r^{obs}_{BAO,i}-r^{th}_{BAO,i}\right]^2}{\sigma_i^2}~,
    \end{equation}
    where $r_{BAO}$ is equivalent to $\frac{r_s(z_d)}{D_V}$ and $\sigma_i$ represents the uncertainty in the measurements for each data point $i$. From now on, this data will be known as the \gls{BAO} data.
    
    \item \textbf{Dark Energy Spectroscopic Instrument (\gls{DESI}):} The \gls{DESI} 2024 dataset represents 2D \gls{BAO} data collected by using the Dark Energy Spectroscopic Instrument. It includes information from bright galaxies, a combination of luminous red galaxies (\gls{LRG}s) and emission-line galaxies (\gls{ELG}s), quasars, and the Lyman-$\alpha$Forest \cite{DESI2024}. \gls{LRG}s are older, massive galaxies, while \gls{ELG}s are younger, actively star-forming galaxies. Quasars are luminous objects fueled by supermassive black holes at the centres of galaxies, and the Lyman-$\alpha$ Forest consists of absorption features in the spectra of distant quasars, which serve as tools for probing the intergalactic medium \cite{RecentDESI}. From this data, \gls{DESI} calculates the values of:
    \begin{eqnarray}
    \frac{D_H(z)}{r_d} &=& \frac{da(1+z)}{r_d}~\label{DH},
    \\
    \frac{D_M(z)}{r_d} &=& \frac{c}{H(z)r_d}~\label{DM},
    \\
    \frac{D_V(z)}{r_d} &=& \frac{(zD_M(z)^2 D_H(z))^{\frac{1}{3}}}{r_d}~\label{DV},
    \end{eqnarray}
    where $r_d$ is the sound horizon at the drag epoch, $D_H$ refers to the Hubble distance, $da$ is the angular distance, $D_M$ represents the comoving angular diameter distance, and $D_V$ is the volume-averaged distance. The \gls{DESI} dataset determines the likelihood function \cite{DESI2024} with the equation:
    \begin{equation}
        \chi^2 = (\textbf{p}_A-\textbf{p}_B)^{T}(Cov_A+Cov_B)^{-1}(\textbf{p}_A-\textbf{p}_B)~,
    \end{equation}
    where $\textbf{p}_A$ and $\textbf{p}_B$ refer to the $\Omega_m$ and $r_dh$ parameters vectors while $Cov_A$ and $Cov_B$ are $2\times 2$ covariances. This dataset will be referred to as the \gls{DESI} data \cite{DESI2024}.

\end{itemize}

For the analysis of the different cosmological scenarios influenced by the dynamical dark energy models, MontePython \cite{MontePython, MontePython2} software was used in conjunction with the software package \gls{CLASS} \cite{CLASS, CLASS_LICENCE}. CLASS is widely used in the research community for its ability to derive constraints on cosmological parameters from data analysis, while MontePython handles the Monte Carlo Markov Chains (\gls{MCMC}). The packages make use of a convergence diagnostic method developed by Gelman and Rubin and support the chosen likelihoods of Planck, Pantheon+SH0ES, Cosmic Chonometers, \gls{BAO} and \gls{DESI}. The software package GetDist \cite{GetDist} was used for the plotting of the triangular plots.

\section{$w$CDM Constraint Analysis} \label{sec:results}

In this section, the key observational findings derived from all the cosmological models analysed using various datasets are presented. The primary focus is on the estimation of $H_0$, as influenced by the underlying cosmological models driven by different dark energy parameterisations. However, the $\sigma_8$ parameter will also be analysed for the different models. For a clearer understanding of $H_0$ and $\sigma_8$ estimations across different datasets and models, we refer to the posterior plots and the tables containing the values obtained for each parameter.

\subsection{Constant parameterisation:}

A combined analysis was carried out for the $w_0$CDM model, with the main observational constraints summarised in Table~\ref{tab:w0CDM_valuesPR3PR4}, and Table~\ref{tab:w0CDM_valuesPR3PR4+CCSN} where each table shows the best-fit, mean and 1$\sigma$ deviation from the mean value for the different data combinations that were taken for the six $\Lambda$CDM parameters and for $w_{0,w0CDM}$, $H_0$, and $\sigma_8$. 

\renewcommand{\thetable}{S\arabic{table}}

\begingroup
\begin{table}
\resizebox{0.8\textwidth}{!}{%
    \makebox[\textwidth][c]{
    \small
    \begin{tabular}{cccccc}
        \hline
        \hline
        Parameters & \multicolumn{2}{c}{\gls{PR3}}& &      \multicolumn{2}{c}{{PR4 }}    \\ 
        \cline{2-3} \cline{5-6}
        &  Best-fit   & Mean&      & Best-fit   & Mean\\    
        \hline
        \textbf{Sampled Parameters}   &     &                                     &  &   &\\
        $\omega_b$ \dotfill  & $0.02247$   & $0.02239^{+0.00015}_{-0.00016}$& & $0.02229$   & $0.02224 \pm{0.00013}$          \\       
        $\omega_{cdm}$\dotfill   &  $0.1191$   &$0.1197 \pm 0.0014$                                   & &$0.1184$  &$0.1188\pm{0.0012}$ \\
        $100\theta_s$\dotfill   &$1.0420$   &$1.0419 ^{+0.00030}_{-0.00031}$& &$1.0418$   & $1.0418 \pm 0.00025$        \\         
        $\ln(10^{10}A_s)$\dotfill   &  $3.054$   & $3.042 ^{+0.016}_{-0.017}$                                 & &$3.0366$   &$3.0383\pm{0.014}$\\
        $n_s$\dotfill   &$0.9654$      &$0.9651^{+0.0043}_{-0.0045}$&     &$0.9691$    & $0.9674^{+0.0042}_{-0.0041}$          \\
        $\tau_{\rm reio}$\dotfill   & $0.0574$ &$0.0592^{+0.0078}_{-0.0087}$& &$0.0593$ &   $0.0576^{+0.0062}_{-0.0063}$          \\
        $w_{0,w0CDM}$\dotfill  & $-2.30$  &$-1.92^{+0.29}_{-0.48}$&  &$-1.24$   &$-1.14 ^{+0.20}_{-0.26}$       \\
        \hline
        \textbf{Derived Parameters}  &     &          \\    
        $H_0$ / km s$^{-1}$Mpc$^{-1}$\dotfill   & $115.45$    &$100.16^{+20.00}_{-10.00}$& &$75.41$    & $72.00^{+8.00}_{-7.00}$  \\     
        $\sigma_8$\dotfill  &  $1.17$ & $1.07^{+0.14}_{-0.06}$ & & $0.871$& $0.840^{+0.071}_{-0.060}$
        \\
        \hline
        \end{tabular}}
    }
    \caption{The values of the constant model obtained from the chosen simulations when only Planck data, \gls{PR3} and \gls{PR4} were used.}
    \label{tab:w0CDM_valuesPR3PR4}           
\end{table}
\endgroup
\renewcommand{\thetable}{S\arabic{table}}

\begingroup
\begin{table}
\resizebox{0.8\textwidth}{!}{%
    \makebox[\textwidth][c]{
    \small
    \begin{tabular}{cccccccccccccc}
        \hline
        \hline
        Parameters & \multicolumn{2}{c}{\gls{PR3}}& & \multicolumn{2}{c}{\gls{PR4} }    & &  \multicolumn{2}{c}{\gls{PR3}}& &  \multicolumn{2}{c}{\gls{PR4} } &\\ 
        & \multicolumn{2}{c}{CC $+$ \gls{SN$+$SH0ES}}&     &\multicolumn{2}{c}{\gls{CC} $+$ \gls{SN$+$SH0ES}}& &\multicolumn{2}{c}{CC $+$ \gls{SN$+$SH0ES} $+$ \gls{DESI}}&     &\multicolumn{2}{c}{\gls{CC} $+$ \gls{SN$+$SH0ES} $+$ \gls{DESI}}\\
        \cline{2-3} \cline{5-6} \cline{8-9} \cline{11-12}
        &  Best-fit   & Mean&      & Best-fit   & Mean & &  Best-fit   & Mean&      & Best-fit   & Mean\\    
        \hline
        \textbf{Sampled Parameters}   &     &                                     &  &   &\\
        $\omega_b$ \dotfill  & $0.02251$   & $0.02252 \pm 0.00015$ &  &$0.02224$ &$0.02235 \pm 0.00012$ & & $0.02258$   & $0.02256\pm 0.00014$ &  &$0.02245$ &$0.02236 \pm 0.00012$        \\       
        $\omega_{cdm}$\dotfill   &  $0.1179$   & $0.1185 \pm 0.0013$                                  &   &$0.11833$  &$0.1176 \pm 0.0011$ & &  $0.1176$   & $0.1180 ^{+0.0011}_{-0.0010}$                                  &   &$0.11704$  &$0.11742^{+0.00092}_{-0.00090}$\\
        $100\theta_s$\dotfill   & $1.04210$   & $1.04207 \pm 0.00029$       & &$1.04180$   &$1.0420 \pm{0.0039}$ && $1.04220$   & $1.04210\pm 0.00029$       & &$1.04200$   &$1.04200 ^{+0.00024}_{-0.00025}$\\         
        $\ln(10^{10}A_s)$\dotfill   &  $3.049$   & $3.050 ^{+0.014}_{-0.015}$                                 &    &$3.041$   &$3.041 \pm 0.014$ &&  $3.056$   & $3.054^{+0.015}_{-0.020}$                                 &    &$3.041$   &$3.040\pm 0.015$\\
        $n_s$\dotfill   & $0.9705$    & $0.9693 ^{+0.0042}_{-0.0047}$    &   &$0.9684$      &$0.9703 \pm 0.0039$ && $0.9747$    & $0.97033^{+0.0040}_{-0.0039}$    &   &$0.9709$      &$0.9707^{+0.0035}_{-0.0036}$       \\
        $\tau_{\rm reio}$\dotfill   & $0.0577$    & $0.0583 ^{+0.0075}_{-0.0073}$    &   &$0.0616$ &$0.0594 \pm 0.0062$   && $0.0623$    & $0.0606^{+0.0073}_{-0.0099}$    &   &$0.0614$ &$0.0597^{+0.0063}_{-0.0067}$              \\
        $w_{0,w0CDM}$\dotfill  &  $-1.034$   & $-1.053 ^{+0.026}_{-0.025}$                                   &   &$-1.062$  &$-1.049\pm 0.024$ & &  $-1.038$   & $-1.049^{+0.026}_{-0.024}$                                   &   &$-1.043$  &$-1.046\pm 0.024$\\
        \hline
        \textbf{Derived Parameters}  &     &          \\    
        $H_0$ / km s$^{-1}$Mpc$^{-1}$\dotfill   &$69.40$    & $69.80 ^{+0.67}_{-0.70}$           &  &$69.74$    &$69.76 \pm 0.66$ &&$69.70$    & $69.86 \pm 0.61$           &  &$69.89$    &$69.76^{+0.63}_{-0.62}$\\
        $\sigma_8$\dotfill  &$0.816$ & $0.823\pm 0.011$&  & $0.823$& $0.818^{+0.011}_{-0.010}$ &&$0.820$ &$0.823^{+0.011}_{-0.013}$ & & $0.813$ & $0.816\pm 0.010$
        \\
        \hline
        \end{tabular}}
    }
    \caption{The values of the constant model obtained from the two Planck data when \gls{CC} and \gls{SN$+$SH0ES} data were added, and when \gls{DESI} was added on top of them.}
    \label{tab:w0CDM_valuesPR3PR4+CCSN}           
\end{table}
\endgroup
\begin{figure}
    \centering
    \includegraphics[width=0.7\linewidth]{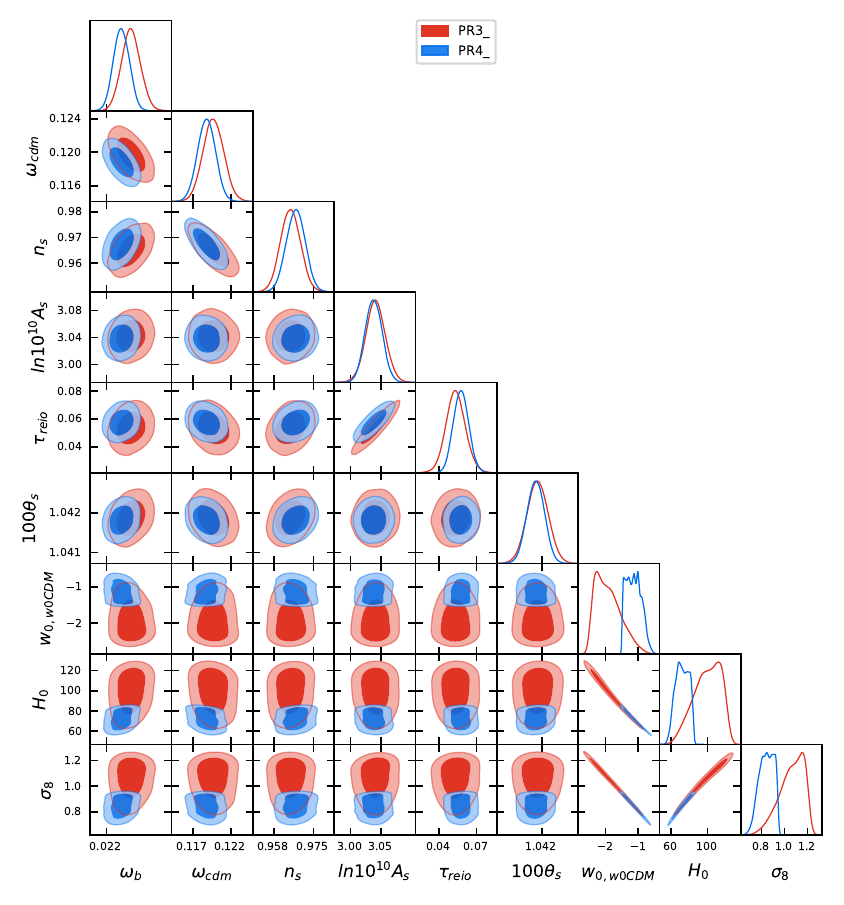}
    \caption{The graph shows the w0CDM model, with \gls{PR3} and \gls{PR4}.}
    \label{fig:w0CDM_ALONE}
\end{figure}
Fig.~\ref{fig:w0CDM_ALONE} shows the 1$\sigma$ and 2$\sigma$ confidence levels in the 2D posteriors of the constant model when using the early time data of \gls{PR3} and \gls{PR4} for comparison. The figure shows uncertainty with regard to the exact value of $w_0$, $H_0$ and $\sigma_8$ in both datasets as degeneracies. When \gls{PR3} was taken, the constant model favours phantom energy and produced higher values of $H_0$ and $\sigma_8$, which go beyond the expected values of past research \cite{Planck2020,Pantheon+,Planck2018,RecentDESI}. It can be concluded that the constant model obtained the smallest posteriors when the \gls{PR4} model was taken, showing that the newer Planck data has less uncertainty regarding the possible values of each parameter. 
\begin{figure}
    \centering
    \includegraphics[width=0.725\linewidth]{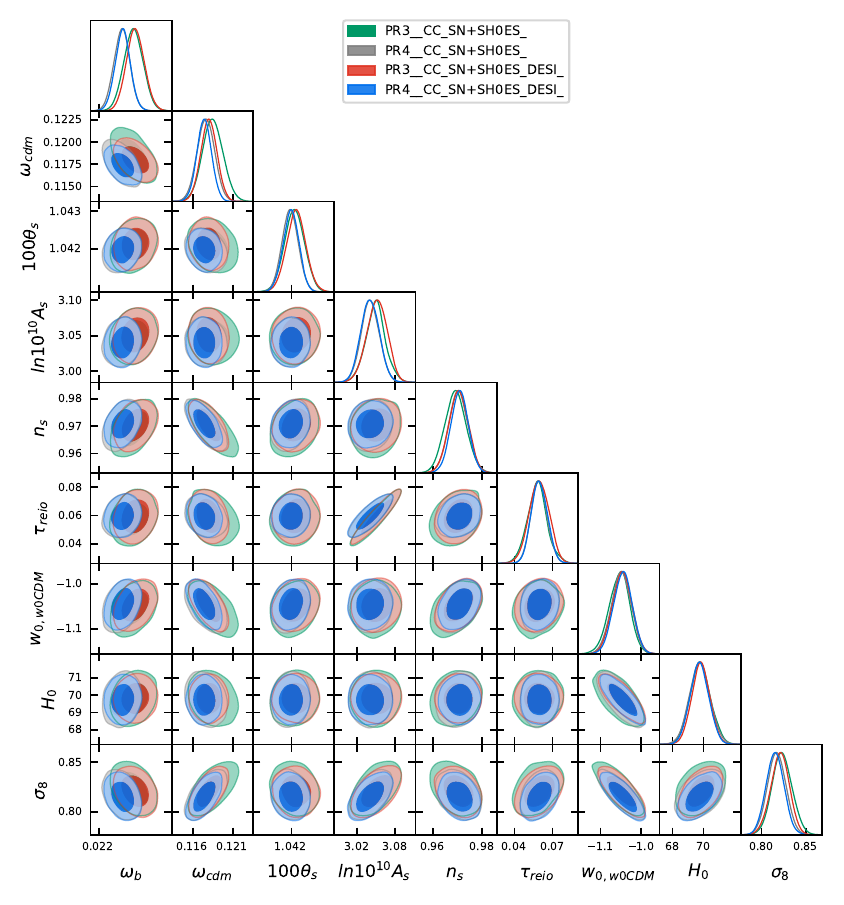}
    \caption{The graph depicts the constant model under different data combinations of \gls{CC} $+$ \gls{SN$+$SH0ES} with \gls{PR3} and \gls{PR4}, and \gls{CC} $+$ \gls{SN$+$SH0ES} $+$ \gls{DESI} with \gls{PR3} and \gls{PR4}.}
    \label{fig:w0CDM_ALL}
\end{figure}
Similarly, Fig.~\ref{fig:w0CDM_ALL} shows the results of the constant model when combining early-time with late-time data. Any degeneracies that were seen in the previous plot, when only early-time data was taken, were removed as soon as late-time data was added to the constant model. The figure illustrates how the addition of the \gls{DESI} data did not affect the results of the model except for constraining the 2D posteriors better. In fact, the constant model obtained the smallest 2D posteriors when \gls{PR4}$+$\gls{CC}$+$\gls{SN$+$SH0ES}$+$\gls{DESI} data combination was used, indicating that this model constrained the parameters the best when considering this data combination. The plot continues to show that the constant model favours a phantom Universe. 

In summary, the constant model favours a phantom Universe, and when using \gls{PR4}, the model is 1$sigma$ away from the $\Lambda$CDM model. Also, both \gls{DESI} and \gls{PR4} show to be better data sets than their predecessors as they constrain the parameters better.

\subsection{Quadratic parameterisation:}

Similar to the constant parametrisation, the quadratic model underwent a series of analyses, with the results summarised in Table~\ref{tab:JBP_valuesPR3PR4}, and Table~\ref{tab:JBP_valuesPR3PR4+CCSN}.
\renewcommand{\thetable}{S\arabic{table}}
\begingroup
\begin{table}
\resizebox{0.8\textwidth}{!}{%
    \makebox[\textwidth][c]{
    \small
    \begin{tabular}{cccccc}
        \hline
        \hline
        Parameters & \multicolumn{2}{c}{\gls{PR3}}& & \multicolumn{2}{c}{\gls{PR4} }    \\ 
        \cline{2-3} \cline{5-6}
        &  Best-fit   & Mean&      & Best-fit   & Mean\\    
        \hline
        \textbf{Sampled Parameters}   &     &                                     &  &   &\\
        $\omega_b$ \dotfill   &$0.02244$   & $0.02238^{+0.00015}_{-0.00016}$ &  & $0.02231$   & $0.02226\pm0.00013$         \\       
        $\omega_{cdm}$\dotfill   &  $0.1177$   & $0.1199\pm 0.0014$                                  &   & $0.11837$   &$0.11870 \pm 0.00012$                                  \\
        $100\theta_s$\dotfill   & $1.04180$   & $1.04190 ^{+0.00031}_{-0.00030}$       & & $1.04190$   & $1.04180 ^{+0.00025}_{-0.00025}$       \\         
        $\ln(10^{10}A_s)$\dotfill   &  $3.044$   & $3.045 \pm 0.016$                                  &    &  $3.047$   & $3.049 ^{+0.014}_{-0.015}$                                \\
        $n_s$\dotfill   &$0.9712$    & $0.9654\pm 0.0046$    &   & $0.9690$    & $0.9676 \pm 0.0041$       \\
        $\tau_{\rm reio}$\dotfill   &$0.0580$    & $0.0543^{+0.0074}_{-0.0081}$    &   & $0.0623$    & $0.0578 ^{+0.0060}_{-0.0066}$  \\
        $w_{0,JBP}$\dotfill  &   $-0.807$   & $-1.235^{+0.004}_{-0.266}$                                   &   & $-1.05$   &$-1.15^{+0.11}_{-0.35}$                                  \\
        $w_{a,JBP}$\dotfill  &  $-1.07$   & $-0.99^{+2.49}_{-0.16}$                                   &   & $-1.00$   & $0.11^{+0.47}_{-0.87}$                                   \\
        \hline
        \textbf{Derived Parameters}  &     &          \\    
        $H_0$ / km s$^{-1}$Mpc$^{-1}$\dotfill   &$ 110.78$    & $79.03^{+4.83}_{-9.14}$           &  & $74.29$    & $73.03^{+8.64}_{-7.52}$   \\
        $\sigma_8$\dotfill &$1.134$&  $0.911^{+0.044}_{-0.078}$&  & $0.867$ &$0.851^{+0.079}_{-0.059}$
        \\
        \hline
        \end{tabular}}
    }
    \caption{The values of the \gls{JBP} model using the two Planck data sets: \gls{PR3} or \gls{PR4}.}
    \label{tab:JBP_valuesPR3PR4}           
\end{table}
\endgroup
\renewcommand{\thetable}{S\arabic{table}}
\begingroup
\begin{table}
\resizebox{0.8\textwidth}{!}{%
    \makebox[\textwidth][c]{
    \small
    \begin{tabular}{cccccccccccccc}
        \hline
        \hline
        Parameters & \multicolumn{2}{c}{\gls{PR3}}& & \multicolumn{2}{c}{\gls{PR4} }    & &  \multicolumn{2}{c}{\gls{PR3}}& &  \multicolumn{2}{c}{\gls{PR4} } &\\ 
        & \multicolumn{2}{c}{CC $+$ \gls{SN$+$SH0ES}}&     &\multicolumn{2}{c}{\gls{CC} $+$ \gls{SN$+$SH0ES}}& &\multicolumn{2}{c}{CC $+$ \gls{SN$+$SH0ES} $+$ \gls{DESI}}&     &\multicolumn{2}{c}{\gls{CC} $+$ \gls{SN$+$SH0ES} $+$ \gls{DESI}}\\
        \cline{2-3} \cline{5-6} \cline{8-9} \cline{11-12}
        &  Best-fit   & Mean&      & Best-fit   & Mean & &  Best-fit   & Mean&      & Best-fit   & Mean\\ 
        \hline
        \textbf{Sampled Parameters}   &     &                                     &  &   &\\
        $\omega_b$ \dotfill  & $0.02250$   & $0.02248^{+0.00014}_{-0.00015}$ &  &$0.02233$ &$0.02232 ^{+0.00013}_{-0.00012}$ && $0.02262$   & $0.02247\pm0.00014$ &  &$0.02238$ &$0.02232 ^{+0.00012}_{-0.00011}$       \\       
        $\omega_{cdm}$\dotfill   &  $0.1192$   & $0.1191 ^{+0.0013}_{-0.0012}$                                  &   &$0.1182$  &$0.1180^{+0.0011}_{-0.0012}$ & &  $0.1188$   & $0.1191 \pm 0.0011$                                  &   &$0.11825$  &$0.11808^{+0.00095}_{-0.00097}$\\
        $100\theta_s$\dotfill   & $1.04190$   & $1.04200^{+0.00028}_{-0.00031}$       & &$1.04200$   &$1.04190 \pm{0.00025}$ && $1.04190$   & $1.04190^{+0.00029}_{-0.00030}$       & &$1.04190$   &$1.04190 ^{+0.00024}_{-0.00023}$\\         
        $\ln(10^{10}A_s)$\dotfill   &  $3.049$   & $3.046^{+0.016}_{-0.018}$                                 &    &$3.025$   &$3.040^{+0.015}_{-0.014}$ &&  $3.044$   & $3.046\pm0.016$                                 &    &$3.034$   &$3.039\pm 0.014$\\
        $n_s$\dotfill   & $0.9653$    & $0.9676 \pm 0.0044$    &   &$0.9670$      &$0.9693\pm 0.0039$      & & $0.9697$    & $0.9674^{+0.0041}_{-0.0040}$    &   &$0.9690$      &$0.9692^{+0.0037}_{-0.0036}$ \\
        $\tau_{\rm reio}$\dotfill   & $0.0550$    & $0.0558^{+0.0078}_{-0.0084}$    &   &$0.0540$ &$0.0590^{+0.0063}_{-0.0065}$  && $0.0528$    & $0.0559^{+0.0076}_{-0.0081}$    &   &$0.0572$ &$0.0585^{+0.0060}_{-0.0064}$        \\
        $w_{0,JBP}$\dotfill  &  $-0.706$   & $-0.790^{+0.090}_{-0.021}$                                   &   &$-0.802$  &$-0.858^{+0.058}_{-0.013}$ &&  $-0.78$   & $-0.81^{+0.10}_{-0.04}$                                   &   &$-0.830$  &$-0.878^{+0.078}_{-0.022}$\\
        $w_{a,JBP}$\dotfill  &  $-2.37$   & $-1.84^{+0.31}_{-0.58}$                                   &   &$-1.63$  &$-1.34^{+0.25}_{-0.42}$ &&  $-1.75$   & $-1.60^{+0.39}_{-0.58}$                                   &   &$-1.41$  &$-1.12^{+0.29}_{-0.47}$\\
        \hline
        \textbf{Derived Parameters}  &     &          \\    
        $H_0$ / km s$^{-1}$Mpc$^{-1}$\dotfill   &$ 70.10$    & $70.24 ^{+0.68}_{-0.69}$           &  &$69.81$    &$70.18\pm 0.68$ &&$ 69.68$    & $69.80 \pm 0.59$           &  &$69.67$    &$69.75^{+0.60}_{-0.63}$\\
        $\sigma_8$\dotfill &$0.834$&  $0.833\pm 0.012$ & &$0.817$& $0.826\pm 0.011$ && $0.825$&  $0.830^{+0.012}_{-0.011}$& &$0.820$& $0.822\pm0.010$
        \\
        \hline
        \end{tabular}}
    }
    \caption{The values obtained from the \gls{JBP} model when using \gls{PR3} and \gls{PR4} with \gls{CC} and \gls{SN$+$SH0ES} data and with \gls{CC}$+$\gls{SN$+$SH0ES}$+$\gls{DESI}.}
    \label{tab:JBP_valuesPR3PR4+CCSN}           
\end{table}
\endgroup
\begin{figure}
    \centering
    \includegraphics[width=0.7\linewidth]{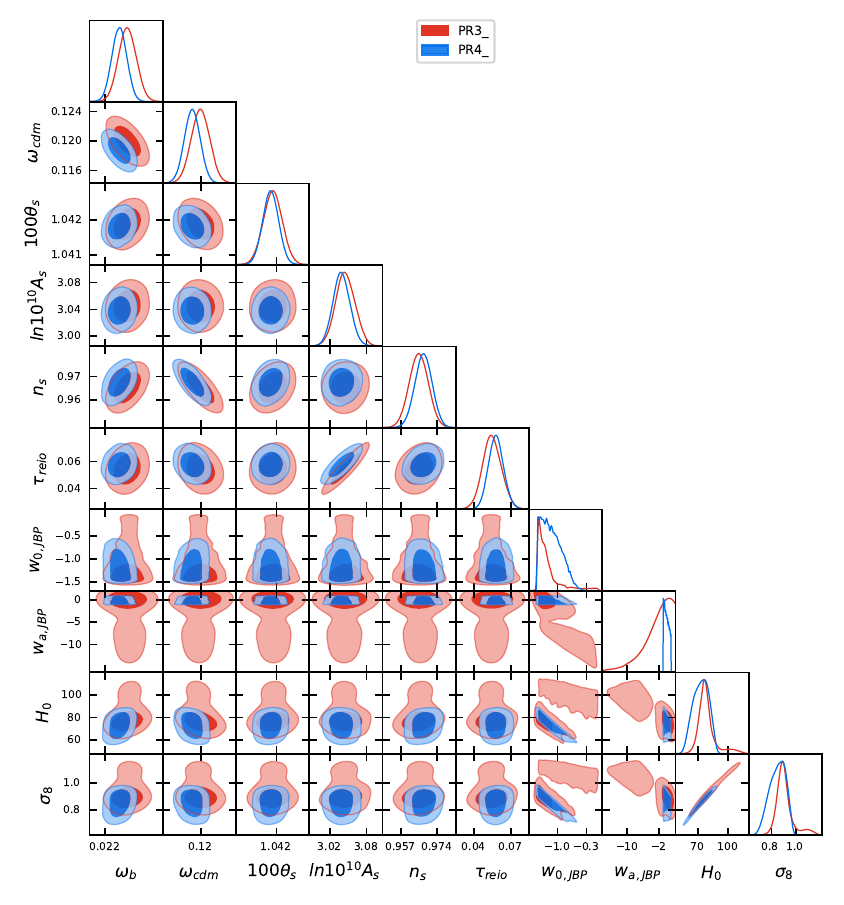}
    \caption{A graph showing the results of the quadratic parameterisation when using \gls{PR3}, and when considering the \gls{PR4} data.}
    \label{fig:JBP_ALONE}
\end{figure}
Fig.~\ref{fig:JBP_ALONE} presents the quadratic parameterisation when the two Planck data were taken separately. From the figure, degeneracies in the quadratic parameterisation, can be spotted with the $w_{0,JBP}$, $w_{a,JBP}$, $H_0$, and $\sigma_8$ parameters when the \gls{PR3} data was considered, showing other possible values that the parameters can obtain apart from the highest peak shown in the 1D posteriors. When comparing the two data sets, it can be seen that the Quadratic model constrains the parameters the best when the latest Planck data is used. When only early-time data is taken, the quadratic model suggests a phantom Universe when the \gls{PR3} or \gls{PR4} data were considered.
\begin{figure}
    \centering
    \includegraphics[width=0.725\linewidth]{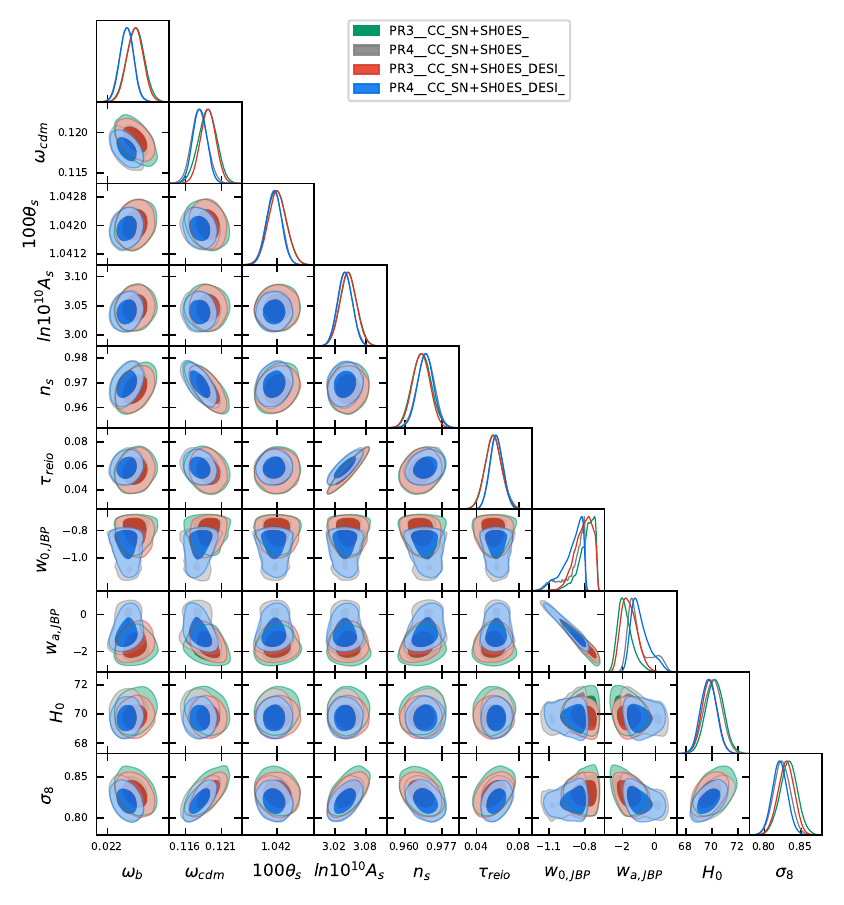}
    \caption{A graph showing the quadratic parameterisation when combinations of early-time and late-time data were considered.}
    \label{fig:JBP_ALL}
\end{figure}
Fig.~\ref{fig:JBP_ALL} displays the parameterisation model, taking combinations of early-time with late-time data. This figure shows that the quadratic model obtained slight degeneracies with the $w_{0,JBP}$ and $w_{a,JBP}$ parameters when the \gls{PR4} $+$ \gls{CC} $+$ \gls{SN$+$SH0ES} data combination was taken, while when the other data combinations were taken, the model did not show any degeneracies. Similar to when late-time data combinations are used, the quadratic model prefers a quintessence Universe at low redshifts, but then favours a Phantom Universe at high redshifts. The smallest posteriors were obtained through the \gls{PR4} $+$ \gls{CC} $+$ \gls{SN$+$SH0ES} $+$ \gls{DESI} data combination, implying that \gls{DESI} and \gls{PR4} help constrain the quadratic model the best. 

In conclusion, when combinations of early-time with late-time data were used, the quadratic model favours a quintessence dark energy at low redshifts, but then changes to a phantom dark energy at high redshifts when late-time data is used. When only early-time data was considered, the model favoured a phantom Universe. The model seems to perform the best when \gls{DESI} and \gls{PR4} data were used, showing that they are more accurate than their predecessor.

\subsection{Logarithmic parameterisation:}

Following the trend of the previous two models, a combination of analyses was performed on the Logarithmic model, and in Table~\ref{tab:GE_valuesPR3PR4}, and Table~\ref{tab:GE_valuesPR3PR4+CCSN} contain the summary of the observational constraints for the Logarithmic model when combination of early-time and early-time with late-time data were taken. 
\renewcommand{\thetable}{S\arabic{table}}
\begingroup
\begin{table}
\resizebox{0.8\textwidth}{!}{%
    \makebox[\textwidth][c]{
    \small
    \begin{tabular}{cccccc}
        \hline
        \hline
        Parameters & \multicolumn{2}{c}{\gls{PR3}}& & \multicolumn{2}{c}{\gls{PR4} }    \\ 
        \cline{2-3} \cline{5-6}
        &  Best-fit   & Mean&      & Best-fit   & Mean\\    
        \hline
        \textbf{Sampled Parameters}   &     &                                     &  &   &\\
        $\omega_b$ \dotfill  & $0.02233$   & $0.02238^{+0.00015}_{-0.00016}$ &  & $0.02231$   & $0.02223\pm{0.00013}$         \\       
        $\omega_{cdm}$\dotfill   &  $0.1203$   & $0.1200\pm 0.0014$                                  &   &  $0.1179$   &$0.1188 \pm 0.0012$                                   \\
        $100\theta_s$\dotfill   & $1.0421$   & $1.0419 \pm 0.00030$       & & $1.0418$   & $1.0418 \pm 0.00025$      \\         
        $\ln(10^{10}A_s)$\dotfill   &  $3.052$   & $3.045\pm 0.016$                                 &    &  $3.034$   & $3.039 \pm 0.014$                                 \\
        $n_s$\dotfill   & $0.9659$    & $0.9653^{+0.0045}_{-0.0046}$    &   & $0.9681$    & $0.9673 \pm{0.0040}$  \\
        $\tau_{\rm reio}$\dotfill   & $0.0578$    & $0.0544^{+0.0076}_{-0.0081}$    &   & $0.0571$    & $0.0577 \pm 0.0062$        \\
        $w_{0,GE}$\dotfill  &  $-1.14$   & $-1.03^{+0.05}_{-0.17}$                                   &   &  $-1.09$   &$-0.99 \pm 0.12$           \\
        $w_{a,GE}$\dotfill  &  $-0.07$   & $-0.18^{+0.07}_{-0.11}$                                   &   &  $-0.014$   & $-0.077 \pm 0.043$                                   \\
        \hline
        \textbf{Derived Parameters}  &     &          \\    
        $H_0$ / km s$^{-1}$Mpc$^{-1}$\dotfill   &$ 72.54$    & $70.77^{+5.02}_{-2.58}$           &  & $71.15$    & $67.62^{+5.02}_{-3.06}$   \\
        $\sigma_8$\dotfill &$0.864$& $0.843^{+0.046}_{-0.022}$  & &$0.829$& $0.806^{+0.046}_{-0.026}$
        \\
        \hline
        \end{tabular}}
    }
    \caption{The values of the \gls{GE} model when only early-time data of \gls{PR3} and \gls{PR4} were used.}
    \label{tab:GE_valuesPR3PR4}           
\end{table}
\endgroup
\renewcommand{\thetable}{S\arabic{table}}
\begingroup
\begin{table}
\makebox[\textwidth][c]{
\resizebox{0.8\textwidth}{!}{%
    \small
    \begin{tabular}{cccccccccccccc}
        \hline
        \hline
        Parameters & \multicolumn{2}{c}{\gls{PR3}}& & \multicolumn{2}{c}{\gls{PR4} }    & &  \multicolumn{2}{c}{\gls{PR3}}& &  \multicolumn{2}{c}{\gls{PR4} } &\\ 
        & \multicolumn{2}{c}{CC $+$ \gls{SN$+$SH0ES}}&     &\multicolumn{2}{c}{\gls{CC} $+$ \gls{SN$+$SH0ES}}& &\multicolumn{2}{c}{CC $+$ \gls{SN$+$SH0ES} $+$ \gls{DESI}}&     &\multicolumn{2}{c}{\gls{CC} $+$ \gls{SN$+$SH0ES} $+$ \gls{DESI}}\\
        \cline{2-3} \cline{5-6} \cline{8-9} \cline{11-12}
        &  Best-fit   & Mean&      & Best-fit   & Mean & &  Best-fit   & Mean&      & Best-fit   & Mean\\ 
        \hline
        \textbf{Sampled Parameters}   &     &                                     &  &   &\\
        $\omega_b$ \dotfill  & $0.02248$   & $0.02251\pm 0.00015$ &  &$0.022426$ &$0.02235 \pm 0.00012$  && $0.02248$   & $0.02251^{+0.00015}_{-0.00016}$ &  &$0.02237$ &$0.02234 \pm 0.00012$      \\       
        $\omega_{cdm}$\dotfill   &  $0.1182$   & $0.1186 \pm 0.0013$                                  &   &$0.11743$  &$0.1176 \pm 0.0011$ & &  $0.1182$   & $0.1186 ^{+0.0013}_{-0.0014}$                                  &   &$0.11734$  &$0.11776^{+0.00091}_{-0.00096}$\\
        $100\theta_s$\dotfill   & $1.0419$   & $1.04205\pm 0.00029$       & &$1.0420$   &$1.04195 \pm{0.00024}$&& $1.04190$   & $1.04210\pm 0.00030$       & &$1.04210$   &$1.04190 \pm{0.00024}$\\         
        $\ln(10^{10}A_s)$\dotfill   &  $3.055$   & $3.046\pm 0.016$                                 &    &$3.0464$   &$3.040\pm 0.014$ & &  $3.055$   & $3.048\pm 0.016$                                 &    &$3.025$   &$3.041^{+0.015}_{-0.014}$\\
        $n_s$\dotfill   & $0.9678$    & $0.9688\pm 0.0043$    &   &$0.9704$      &$0.9704\pm 0.0039$    && $0.9678$    & $0.9688^{+0.0044}_{-0.0045}$    &   &$0.9707$      &$0.9700\pm 0.0035$    \\
        $\tau_{\rm reio}$\dotfill   & $0.0630$    & $0.0565^{+0.0074}_{-0.0082}$    &   &$0.0588$ &$0.0593\pm 0.0063$     && $0.0622$    & $0.0574^{+0.0078}_{-0.0081}$    &   &$0.0525$ &$0.0592^{+0.0059}_{-0.0069}$      \\
        $w_{0,GE}$\dotfill  &  $-0.932$   & $-0.960^{+0.034}_{-0.082}$                                   &   &$-0.978$  &$-1.004^{+0.029}_{-0.026}$ & &  $-0.932$   & $-0.970^{+0.034}_{-0.030}$                                   &   &$-0.979$  &$-1.000^{+0.028}_{-0.025}$\\
        $w_{a,GE}$\dotfill  &  $-0.274$   & $-0.246^{+0.025}_{-0.056}$                                   &   &$-1.9132$  &$-0.149^{+0.024}_{-0.052}$ &&  $-0.273$   & $-0.245^{+0.011}_{-0.055}$                                   &   &$-0.195$  &$-0.146^{+0.013}_{-0.054}$\\
        \hline
        \textbf{Derived Parameters}  &     &          \\    
        $H_0$ / km s$^{-1}$Mpc$^{-1}$\dotfill   &$ 69.83$    & $70.22 \pm 0.68$           &  &$57.22$    &$63.60\pm 4.80$ &&$ 69.83$    & $70.20 ^{+0.66}_{-0.68}$           &  &$56.17$    &$64.10^{+5.80}_{-3.90}$\\
        $\sigma_8$\dotfill &$0.828$ &$0.830\pm 0.012$   & & $0.713$& $0.767^{+0.052}_{-0.040}$ &&$0.828$& $0.830\pm 0.012$   & &$0.696$& $0.773^{0.049}_{-0.032}$
        \\
        \hline
        \end{tabular}}
    }
    \caption{The values of the logarithmic model when using \gls{PR3} and \gls{PR4} with \gls{CC} and \gls{SN$+$SH0ES} data. Then when \gls{DESI} was added to the data combinations.}
    \label{tab:GE_valuesPR3PR4+CCSN}  
\end{table}
Fig.~\ref{fig:GE_ALONE} shows the Logarithmic model when only early-time data of \gls{PR3} and \gls{PR4} were taken. The figure shows that when the Logarithmic model was considered, the model had degeneracies with obtaining the values of $w_{0,GE}$, $w_{a,GE}$, and $H_0$ when \gls{PR3} was taken and $w_{a,GE}$ when \gls{PR4} was included, as degeneracies can be seen. When \gls{PR3} was taken, this model seems to favour a phantom Universe, while this was not the case when \gls{PR4} was considered, as it suggests a quintessence dark energy at current time, but then changes to a phantom Universe at higher redshifts. Unlike in the previous models, the Logarithmic parametrisation obtained smaller posteriors with the six $\Lambda$CDM parameters and $w_{a,GE}$ parameter when \gls{PR4} was used, but obtained smaller posteriors with $w_{0,GE}$, $H_0$, and $\sigma_8$ when \gls{PR3} was used.  
\begin{figure}
    \centering
    \includegraphics[width=0.7\linewidth]{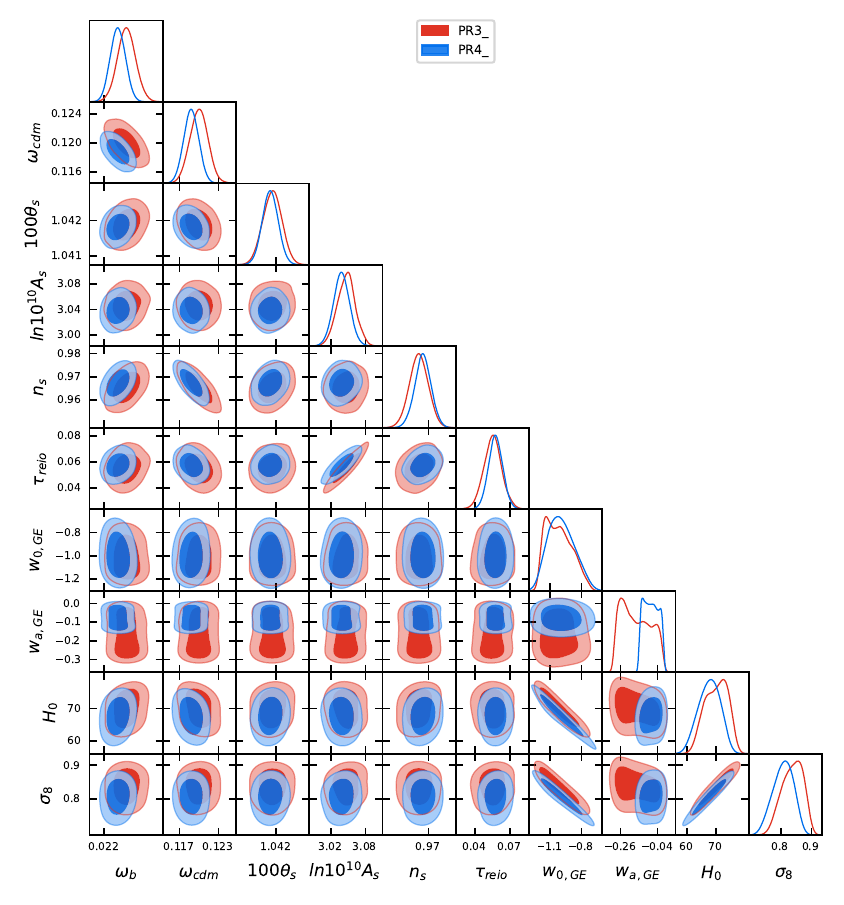}
    \caption{The Logarithmic parameterisation is depicted using observational data, when using \gls{PR3}, and \gls{PR4}.}
    \label{fig:GE_ALONE}
\end{figure}
Fig.~\ref{fig:GE_ALL} depicts the Logarithmic model when combinations of early-time with late-time were taken. The figure demonstrates that the addition of the late-time data was enough to remove any degeneracies with the parameter $w_{0,GE}$ that were seen in the previous graph. It is also seen that when \gls{PR3} was considered, the Logarithmic model had no degeneracies with any of the parameters. This is not the case for when \gls{PR4} was taken, as degeneracies with the $H_0$ and $\sigma_8$ parameters are found. This figure also shows that the addition of the \gls{DESI} data did not affect this model. Unlike when only Planck data was considered, when late-time data was added, the Logarithmic parameterisation shows a preference towards a quintessence-like Universe at current times. However, as the redshift increases, the Logarithmic model prefers a phantom Universe. It is important to note that the Logarithmic model obtained a higher value of $\sigma_8$ than what was found in past research \cite{Planck2018,Pantheon+}. From this figure, it can be noted that the Logarithmic parameterisation achieves the smallest 2D posteriors \gls{PR4}$+$\gls{CC}$+$\gls{SN$+$SH0ES}$+$\gls{DESI} data combination for the six $\Lambda$CDM parameters and for the $w_{0,GE}$ parameter, while for the other parameters the logarithmic model obtained the smallest 2D posteriors when \gls{PR3}$+$\gls{CC}$+$\gls{SN$+$SH0ES}$+$\gls{DESI} data combination was used.
\begin{figure}
    \centering
    \includegraphics[width=0.725\linewidth]{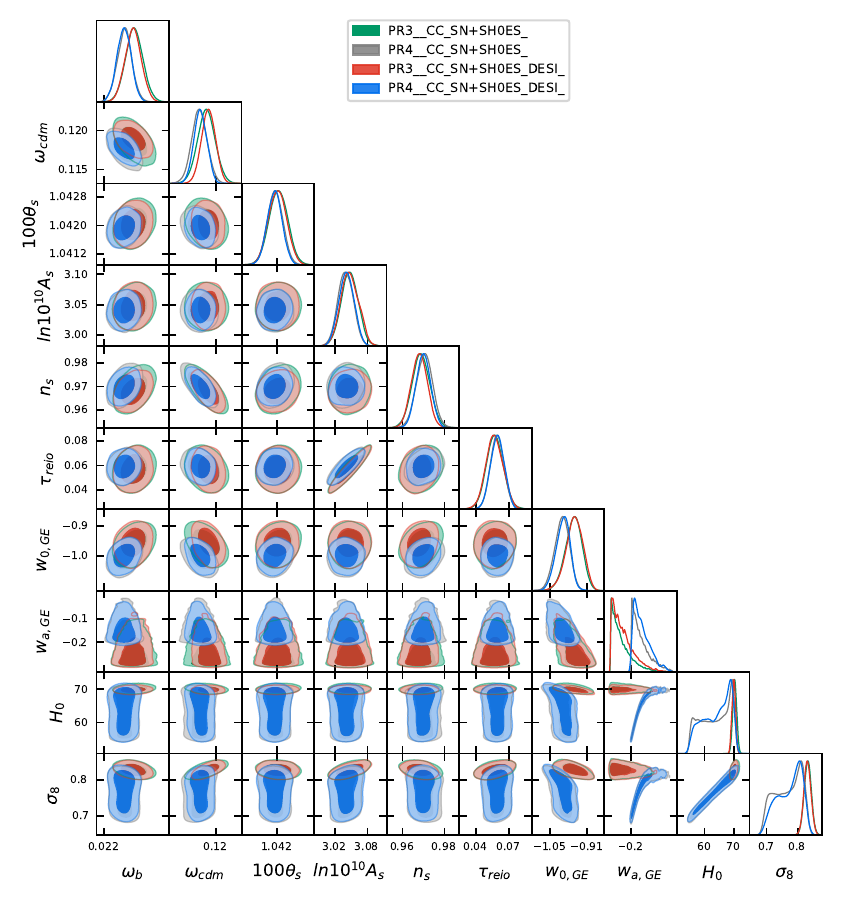}
    \caption{The graph shows the logarithmic model using late-time data combinations alongside \gls{PR3} and \gls{PR4}.}
    \label{fig:GE_ALL}
\end{figure}
In conclusion, the Logarithmic model only obtained values of $w_{0,GE}$ and $w_{a,GE}$ that are 1$\sigma$ away from the standard model when \gls{PR3}$+$\gls{CC}$+$\gls{SN$+$SH0ES} data combination was taken. The Logarithmic model opts for a phantom Universe when \gls{PR3} by itself was considered, and when \gls{PR4} with late-time data combinations were taken. For the other data combinations, the Logarithmic parameterisation model favours a quintessence dark energy at low redshifts and then favours a phantom dark energy at high redshifts. The \gls{DESI} data has been shown to constrain the parameters better when it is used on this model. On the other hand, unlike in the previous parameterisation models, the \gls{PR4} data was a better data set when constraining the six $\Lambda$CDM parameters and the $w_{0,GE}$ parameter. However, when it came to the parameters of $w_{a,GE}$, $H_0$, and $\sigma_8$, the previously released Planck data performed better than the newer Planck data.

\subsection{Oscillatory parameterisation:}

The analysis of the Oscillatory parameterisation was conducted, following the trend of previous models, Table~\ref{tab:OSCILL_valuesPR3PR4}, and Table~\ref{tab:OSCILL_valuesPR3PR4+CCSN} present the results achieved from the Oscillatory model when using the same data combinations as used in the previous models. 
\renewcommand{\thetable}{S\arabic{table}}
\begingroup
\begin{table}
\resizebox{0.8\textwidth}{!}{%
    \makebox[\textwidth][c]{
    \small
    \begin{tabular}{cccccc}
        \hline
        \hline
        Parameters & \multicolumn{2}{c}{\gls{PR3}}& & \multicolumn{2}{c}{\gls{PR4} }    \\ 
        \cline{2-3} \cline{5-6}
        &  Best-fit   & Mean&      & Best-fit   & Mean\\    
        \hline
        \textbf{Sampled Parameters}   &     &                                     &  &   &\\
        $\omega_b$ \dotfill  &$0.02249$   & $0.02240^{+0.00015}_{-0.00016}$ &  & $0.02222$   & $0.02225\pm 0.00013$  \\       
        $\omega_{cdm}$\dotfill   & $0.1191$   & $0.1198\pm 0.0014$                                  &   &  $0.1192$   &$0.1187\pm 0.0012$        \\
        $100\theta_s$\dotfill   & $1.04170$   & $1.04190 \pm 0.00030$       & & $1.04180$   & $1.04180\pm 0.00025$     \\         
        $\ln(10^{10}A_s)$\dotfill   &$3.040$   & $3.046\pm 0.018$                                 &    &  $3.036$   & $3.038^{+0.014}_{-0.015} $                             \\
        $n_s$\dotfill   &$0.9678$    & $0.9658^{+0.0044}_{-0.0046}$    &   & $0.9694$    & $0.9677^{+0.0041}_{-0.00040}$ \\
        $\tau_{\rm reio}$\dotfill   & $0.0520$    & $0.0552^{+0.0085}_{-0.0087}$    &   & $0.0580$    & $0.0576^{+0.0060}_{-0.0064}$ \\
        $w_{0,OSCILL}$\dotfill  &$-1.39$   & $-1.17^{+0.07}_{-0.33}$                                   &   &  $-0.83$   &$-1.30^{+0.34}_{-0.39} $   \\
        $w_{a,OSCILL}$\dotfill  &  $0.71$   & $1.25^{+0.08}_{-1.10}$                                   &   &   $1.52$   & $0.18^{+1.30}_{-0.96}$   \\
        \hline
        \textbf{Derived Parameters}  &     &          \\    
        $H_0$ / km s$^{-1}$Mpc$^{-1}$\dotfill   &$ 83.41$    & $77.39^{+9.98}_{-8.03}$           &  & $69.09$    & $78.41^{+10.32}_{-14.85}          $\\
        $\sigma_8$\dotfill &$0.942$ &$0.901^{+0.082}_{-0.067}$ & &$0.83$ &$0.89^{+0.29}_{-0.27}$
        \\
        \hline
        \end{tabular}}
    }
    \caption{The values obtained from oscillatory reparametrisation model using either \gls{PR3} or \gls{PR4}.}
    \label{tab:OSCILL_valuesPR3PR4}           
\end{table}
\endgroup
\renewcommand{\thetable}{S\arabic{table}}
\begingroup
\begin{table}
\resizebox{0.8\textwidth}{!}{%
    \makebox[\textwidth][c]{
    \small
    \begin{tabular}{cccccccccccccc}
        \hline
        \hline
        Parameters & \multicolumn{2}{c}{\gls{PR3}}& & \multicolumn{2}{c}{\gls{PR4} }    & &  \multicolumn{2}{c}{\gls{PR3}}& &  \multicolumn{2}{c}{\gls{PR4} } &\\ 
        & \multicolumn{2}{c}{CC $+$ \gls{SN$+$SH0ES}}&     &\multicolumn{2}{c}{\gls{CC} $+$ \gls{SN$+$SH0ES}}& &\multicolumn{2}{c}{CC $+$ \gls{SN$+$SH0ES} $+$ \gls{DESI}}&     &\multicolumn{2}{c}{\gls{CC} $+$ \gls{SN$+$SH0ES} $+$ \gls{DESI}}\\
        \cline{2-3} \cline{5-6} \cline{8-9} \cline{11-12}
        &  Best-fit   & Mean&      & Best-fit   & Mean & &  Best-fit   & Mean&      & Best-fit   & Mean\\ 
        \hline
        \textbf{Sampled Parameters}   &     &                                     &  &   &\\
        $\omega_b$ \dotfill  & $0.02247$   & $0.02251^{+0.00016}_{-0.00015}$ &  &$0.02240$ &$0.02234 \pm 0.00012$  && $0.02251$   & $0.02250 \pm 0.00014$ &  &$0.02219$ &$0.02226 \pm 0.00013$         \\       
        $\omega_{cdm}$\dotfill   &  $0.1184$   & $0.1188 ^{+0.0013}_{-0.0011}$                                  &   &$0.1173$  &$0.1178\pm 0.0011$ & &  $0.1185$   & $0.1187 ^{+0.0010}_{-0.0011}$                                  &   &$0.11888$  &$0.11895^{+0.00094}_{-0.00099}$\\
        $100\theta_s$\dotfill   & $1.04200$   & $1.04200^{+0.00028}_{-0.00030}$       & &$1.04190$   &$1.04190 ^{+0.00025}_{-0.00026}$ & & $1.04180$   & $1.04200 \pm 0.00029$       & &$1.04190$   &$1.04180 \pm 0.00025$\\         
        $\ln(10^{10}A_s)$\dotfill   &  $3.056$   & $3.047^{+0.016}_{-0.018}$                                 &    &$3.050$   &$3.039\pm 0.014$ &  &  $3.046$   & $3.046^{+0.015}_{-0.017}$                                 &    &$3.055$   &$3.037 \pm 0.014$\\
        $n_s$\dotfill   & $0.9701$    & $0.9686^{+0.0041}_{-0.0040}$    &   &$0.9721$      &$0.9700^{+0.0037}_{-0.0041}$      && $0.9686$    & $0.9685^{+0.0040}_{-0.0038}$    &   &$0.9667$      &$0.9669\pm 0.0035$    \\
        $\tau_{\rm reio}$\dotfill   & $0.0626$    & $0.0566^{+0.0074}_{-0.0088}$    &   &$0.0642$ &$0.0588^{+0.0062}_{-0.0064}$     && $0.0563$    & $0.0563^{+0.0075}_{-0.0085}$    &   &$0.0600$ &$0.0568^{+0.0059}_{-0.0061}$           \\
        $w_{0,OSCILL}$\dotfill  &  $-0.878$   & $-0.904^{+0.042}_{-0.023}$                                   &   &$-0.836$  &$-0.863^{+0.039}_{-0.050}$ &&  $-0.980$   & $-0.997^{+0.030}_{-0.027}$                                   &   &$-0.938$  &$-0.910^{+0.042}_{-0.043}$\\
        $w_{a,OSCILL}$\dotfill  &  $1.28$   & $1.23^{+0.27}_{-0.70}$                                   &   &$1.83$  &$1.51^{+0.43}_{-0.20}$ &&  $-0.48$   & $-0.38^{+0.12}_{-0.02}$                                   &   &$0.92$  &$1.08^{+0.28}_{-0.33}$\\
        \hline
        \textbf{Derived Parameters}  &     &          \\    
        $H_0$ / km s$^{-1}$Mpc$^{-1}$\dotfill   &$ 70.30$    & $70.74 ^{+0.66}_{-0.74}$           &  &$71.45$    &$70.80^{+0.70}_{-0.71}$ &&$ 69.92$    & $69.87 ^{+0.61}_{-0.62}$           &  &$70.07$    &$69.85^{+0.60}_{-0.59}$\\
        $\sigma_8$\dotfill  &$0.837$ &$0.838^{+0.011}_{-0.012}$ & &$0.841$ &$0.833 \pm 0.010$ &&$ 0.826$    & $0.827^{+0.011}_{-0.012}$           &  &$0.836$    &$0.831 \pm 0.010$
        \\
        \hline
        \end{tabular}}
    }
    \caption{The values that were achieved from the Oscillatory parameter when using \gls{CC} and \gls{SN$+$SH0ES}, either with \gls{PR3} or \gls{PR4} for the observational data. Then adding \gls{DESI} to the two data combiantions.}
    \label{tab:OSCILL_valuesPR3PR4+CCSN}           
\end{table}
\endgroup
\begin{figure}
    \centering
    \includegraphics[width=0.7\linewidth]{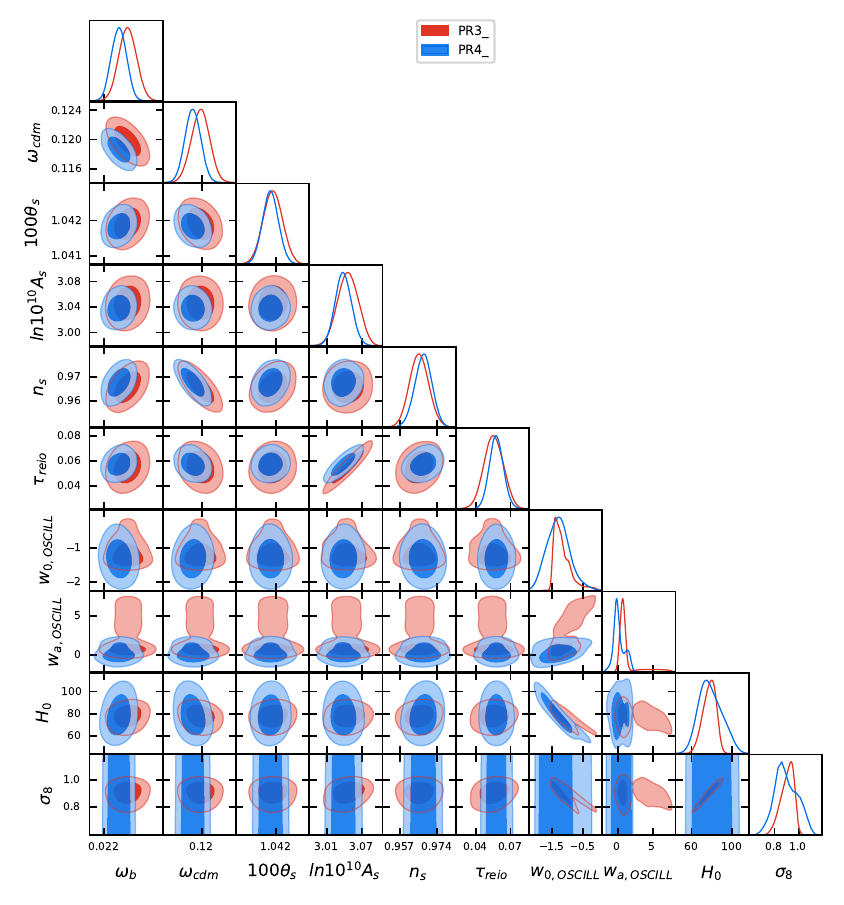}
    \caption{The graph illustrates the two \gls{CMB} data sets, unaffected by late-time data for the Oscillatory model.}
    \label{fig:OSCILL_ALONE}
\end{figure}
Fig.~\ref{fig:OSCILL_ALONE} demonstrates the results of the Oscillatory model when Planck data by itself is incorporated. A complete degeneracy with the $\sigma_8$ parameter is seen in this model when the latest Planck data was used. The Oscillatory parameterisation did not have a complete degeneracy when \gls{PR3} was considered. The Oscillatory parameterisation assumes a phantom Universe when only Planck data was taken. The values of $H_0$ and $\sigma_8$ are higher than the expected values of past research \cite{Planck2018,DESI2024,Pantheon+}. From the 2D posteriors, it is concluded that for the six $\Lambda$CDM parameters and the $w_{a,OSCILL}$ parameter, the oscillatory model achieved smaller posteriors with the \gls{PR4} data. As for the rest of the parameters, it can be seen that this model produced smaller posteriors when the previously released Planck data, \gls{PR3}, was used.
\begin{figure}
    \centering
    \includegraphics[width=0.725\linewidth]{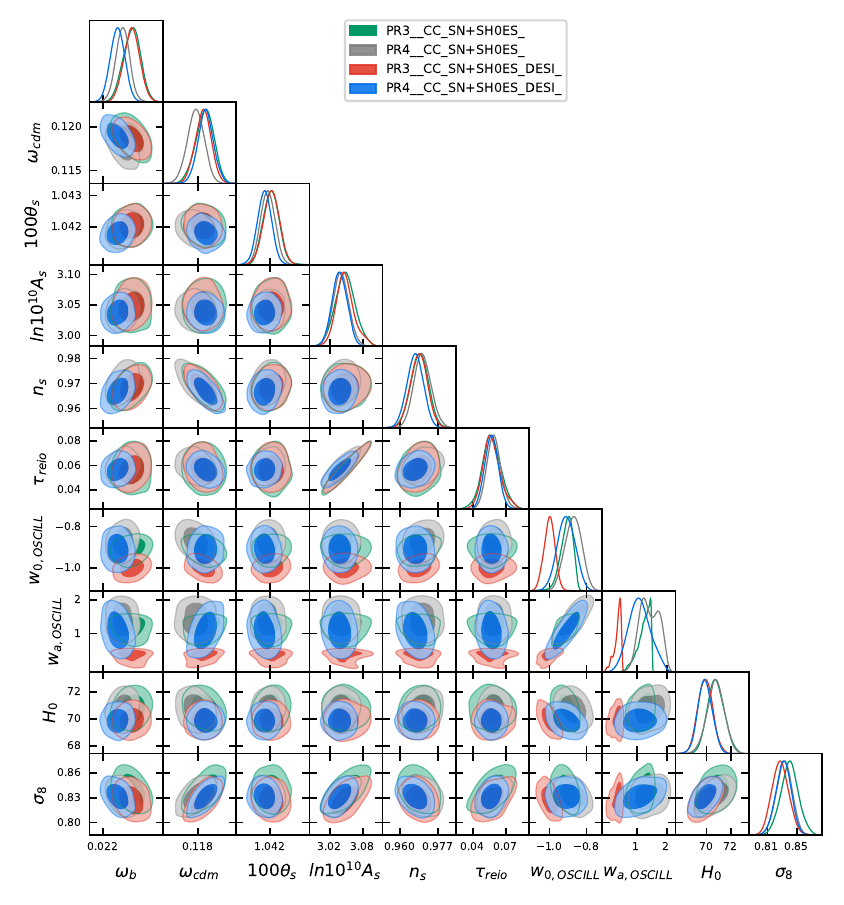}
    \caption{The graph presents four overlapping corner plots of the Oscillatory model, incorporating two Planck data sets combined with various late-time data.}
    \label{fig:OSCILL_ALL}
\end{figure}
Fig.~\ref{fig:OSCILL_ALL} shows the results of this model when combinations of early-time with late-time data were taken. The addition of the late-time data was enough to constrain the complete degeneracy that was found when \gls{PR4} data was used. When the late-time data was added, the Oscillatory model turned from obtaining a phantom Universe to a quintessence one at low redshifts, as the late-time data lowered the values of $w_{0,OSCILL}$ to then change to a phantom Universe at high redshifts. From the four data combinations, it is seen that the oscillatory parameterisation \gls{PR4}$+$\gls{CC}$+$\gls{SN$+$SH0ES}$+$\gls{DESI} produced the smallest with the six $\Lambda$ parameters, $H_0$, and $\sigma_8$, while for the $w_{0,OSCILL}$, and $w_{a,OSCILL}$ parameters, the oscillatory model obtained the smallest posteriors when the \gls{PR3}$+$\gls{CC}$+$\gls{SN$+$SH0ES}$+$\gls{DESI} data combination was considered.

In summary, the Oscillatory parameterisation is 1$\sigma$ away from the $\Lambda$CDM model when \gls{PR4} data combination was taken. This model produced a phantom Universe when only Planck data was used, but then produced a quintessence Universe at low redshifts and a phantom Universe at high redshifts when late-time data was used and when late-time with early-time data was considered. This model showed that the \gls{DESI} data showed to constrain the models better when it is added, while \gls{PR4} showed to be a better dataset with the six $\Lambda$CDM parameters, and the $w_{a,OSCILL}$ parameter. The \gls{PR3} showed to constrain the parameters $w_{0,OSCILL}$, $H_0$, and $\sigma_8$.\\

To conclude, this section showed the results that were obtained from four different dark energy equation of state parameters, showcased in tables displaying the mean, 1$\sigma$ deviation, and best-fit, and in corner plots showing the marginalised 1-dimensional posteriors and the 2-dimensional 1$\sigma$ and 2$\sigma$ confidence-level posteriors. The constant model gives a phantom dark energy, the Logarithmic model generally prefers a quintessence dark energy, and the Oscillatory model showed a quintessence dark energy when late-time and late-time with early-time were considered, while when only early-time was taken, the models obtained a phantom dark energy. The tested models showed that generally, \gls{DESI} constrained the parameters better when it was added to the models, while \gls{PR4} showed to be a better dataset than the previously released datasets of \gls{PR3}, respectively. However, for specific models, and for certain parameters, the \gls{PR3} showed to perform better than the newer Planck data.

\section{Analysis of Results} \label{sec:analysis}

In this section, the statistical performance of each parameterisation will be discussed, as well as how the models affect the six $\Lambda$CDM parameters. The Akaike Information Criterion (\gls{AIC}) and Bayesian Information Criterion (\gls{BIC}) statistics were taken as a measure to determine which parameterisation models are statistically better than the $\Lambda$CDM model and which ones are not. 
\renewcommand{\thetable}{S\arabic{table}}
\begingroup
\begin{table}
\resizebox{0.8\textwidth}{!}{%
    \makebox[\textwidth][c]{
    \begin{tabular}{lccccccccc}
    \hline
    \hline
        $\Delta$AIC& CCSN & CCSNBAO & \gls{ALL}& \gls{PR3} & \gls{PR4} & PR3/CCSN & PR4/CCSN & \gls{PR3$/$ALL} & \gls{PR4$/$ALL}\\
        $w$CDM models\\
         \hline
         Constant \dotfill & $0.91$ & $-4.38$ & $-0.65$ & $-4.40$ &$1.80$ &$-4.44$ & $-1.40$ &$-3.24$ &$-1.00$\\
         Quadratic \dotfill & $0.61$ &$-0.38$ &$-3.71$ &$-2.08$ &$3.60$ &$-11.36$ &$-8.40$ &$-9.48$ &$-6.40$\\
         Logarithmic \dotfill & $0.93$ &$4.02$ &$-1.86$ &$1.72$ &$3.00$ &$-11.54$ &$-6.00$ &$-25.52$ &$-4.60$\\
         Oscillatory \dotfill & $0.89$ &$-2.31$ &$-7.48$ &$-0.96$ &$2.67$ &$-17.30$ &$-18.00$ &$-10.62$ &$-8.80$\\
         \hline
        \hline
        $\Delta$BIC& CCSN & CCSNBAO & \gls{ALL}& \gls{PR3} & \gls{PR4} & PR3/CCSN & PR4/CCSN & \gls{PR3$/$ALL} & \gls{PR4$/$ALL}\\
        $w$CDM models\\
         \hline
         Constant \dotfill & $6.37$ & $1.09$ & $4.81$ &$10.04$ &$9.90$ &$3.60$ &$6.96$ &$4.80$ &$7.36$\\
         Quadratic \dotfill & $13.52$ &$1.09$ &$7.22$ &$14.00$ &$20.32$ &$4.72$ &$8.32$ &$6.60$ &$10.32$\\
         Logarithmic \dotfill & $13.84$ &$14.95$ &$9.06$ &$17.80$ &$19.61$ &$4.54$ &$10.72$ &$-9.44$ &$12.12$\\
         Oscillatory \dotfill & $13.80$ &$8.62$ &$3.45$ &$15.12$ &$19.21$ &$-1.23$ &$-2.04$ &$5.46$ &$7.92$\\
         \hline
    \end{tabular}}
    }
    \caption{The table shows the values of the $\Delta$\gls{AIC} and the $\Delta$BIC that were obtained from the four $w$CDM parametrisation model for each data combination, where CCSN is short for \gls{CC}$+$\gls{SN$+$SH0ES}, CCSNBAO refers to \gls{CC}$+$\gls{SN$+$SH0ES}$+$\gls{BAO}, \gls{ALL} is a notation for \gls{CC}$+$\gls{SN$+$SH0ES}$+$DESI, \gls{PR3}/CCSN is short for \gls{PR3}$+$CC$+$SN+SH0ES, \gls{PR4}/CCSN refers to \gls{PR4}$+$\gls{CC}$+$\gls{SN$+$SH0ES}, \gls{PR3$/$ALL} represents PR3$+$\gls{CC}$+$\gls{SN$+$SH0ES}$+$DESI, and \gls{PR4$/$ALL} represents \gls{PR4}$+$\gls{CC}$+$\gls{SN$+$SH0ES}$+$DESI.}
    \label{tab:ALL_AIC&BIC}
\end{table}
\endgroup
At the top of Table~\ref{tab:ALL_AIC&BIC} show the calculated values $\Delta$\gls{AIC}, which is \gls{AIC}$_{model}-$\gls{AIC}$_{\Lambda CDM}$ and the values of the $\Delta$\gls{BIC} which was calculated as \gls{BIC}$_{model}-$\gls{BIC}$_{\Lambda CDM}$, respectively. When the latest Planck data was used, \gls{PR4}, the $\Lambda$CDM model fit the data better than any of the tested models. On the other hand, when the previously released Planck data was considered, all the models, except for the Logarithmic model, fit the data better than the standard model. When only late-time data that included \gls{DESI} was considered, all the models were statistically better than the $\Lambda$CDM model. From the \gls{AIC} statistics, it can be seen that the model that fits the data the best depends on which data combination was picked. When looking at the $\Delta$\gls{BIC}, the only cases where any of the models fit the data better than the standard models were: the Oscillatory parameterisation when the \gls{PR3}$+$\gls{CC}$+$\gls{SN$+$SH0ES} data combination and the \gls{PR4}$+$\gls{CC}$+$\gls{SN$+$SH0ES} were considered; and the Logarithmic model when \gls{PR3}$+$\gls{CC}$+$\gls{SN$+$SH0ES}$+$D\gls{DESI} data combination was used. From the \gls{BIC} statistics, it can be concluded that the model that fit the data the best out of the five chosen parameterisations was the Oscillatory model. 
\begin{figure}
    \centering
    \includegraphics[width=0.675\linewidth]{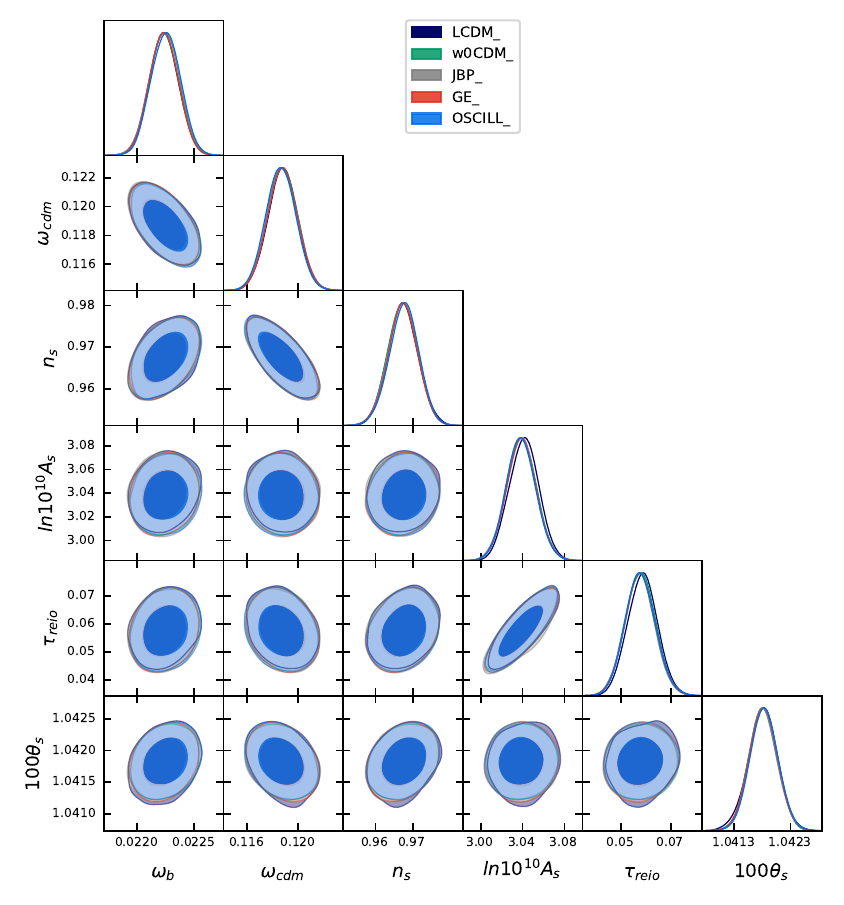}
    \caption{The graph displays five overlapping corner plots, highlighting the six $\Lambda$CDM parameters extracted from the standard model alongside the four $w$CDM models when the \gls{PR4} data set was used.}
    \label{fig:Param_ALONE}
\end{figure}
Fig.~\ref{fig:Param_ALONE} was generated showing the six $\Lambda$CDM parameters of the four $w$CDM models with respect to the standard model. It shows clearly the 1D marginalised posterior distributions and 2D contour plots for all considered dark energy models when using \gls{PR4} alone. There are no notable differences in the 1D and 2D posteriors between each tested model and the standard model. This shows that when early-time data was used, all the models have no effect on the early Universe as the models decribe the same early Universe as the standard model. The constraints on the free cosmological parameters remain largely unaffected by the choice of dark energy parametrisation. 
\begin{figure}
    \centering
    \includegraphics[width=0.675\linewidth]{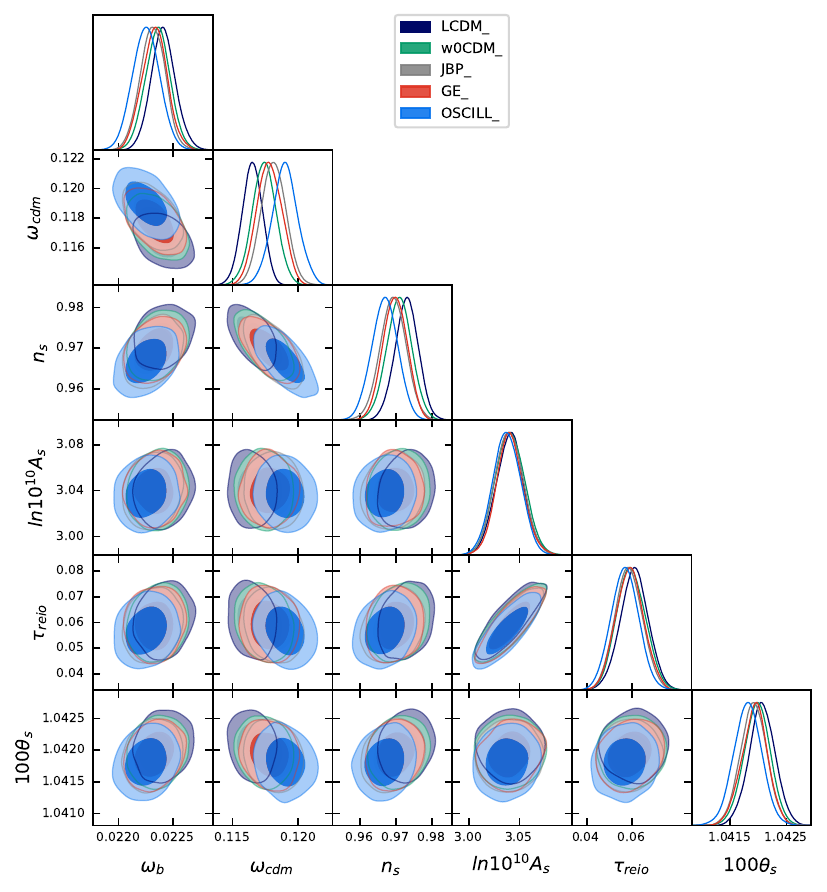}
    \caption{A graph overlapping five corner plots, showing the six $\Lambda$CDM parameters that were achieved from the standard model and the four $w$CDM models that were tested in this project using the \gls{PR4}$+$\gls{CC}$+$\gls{SN$+$SH0ES}$+$\gls{DESI} data combination.}
    \label{fig:PARAM_ALL}
\end{figure}
Fig.~\ref{fig:PARAM_ALL} depicts the 1D marginalised posteriors and the 2D posteriors showing the 1$\sigma$ and 2$\sigma$ confidence levels of the six $\Lambda$CDM parameters for all the models, including the $\Lambda$CDM model when using \gls{PR4} with \gls{CC}, \gls{SN$+$SH0ES}, and \gls{DESI}. When late time was added to the models, the parameters obtained from the chosen models deviated from the standard model. Therefore, when late-time data was included in the $w$CDM models, effect the early Universe as is predicts the six $\Lambda$CDM parameters to be slightly different from the standard model. However, the 2D posteriors show that despite the deviation from the standard model, the models are all $1\sigma$ away from the $\Lambda$CDM model showing that despite this deviation, all the tested models limit to the standard model when it comes to the six $\Lambda$CDM parameters. \\

The vertical coloured bars show the values of $H_0$ of past research: the cyan shows the value from Planck 2018 of $67.4 \pm 0.5 $ km s$^{-1}$Mpc$^{-1}$ \cite{Planck2018}; and the pink shows the value from SH0ES of $73.2 \pm 1.3 $ km s$^{-1}$Mpc$^{-1}$ \cite{SH0ES}. The figure indicates that, when relying solely on late-time data, the $w$CDM models consistently constrained $w_0$, yielding very low uncertainties except for the constant model when \gls{DESI} was incorporated. Regarding the $w_a$ parameter, the logarithmic model achieved the lowest uncertainties and was closest to the standard model's value. On the contrary, the Oscillatory model and the Quadratic parameterisation had the largest uncertainty ranges. As for $H_0$, the reparameterization models, generally attained an uncertainty range comparable to that of the $\Lambda$CDM model, except for when the logarithmic model was taken, as it consistently obtained high uncertainty bars for the $H_0$ parameter. 
\begin{figure}
    \centering
    \includegraphics[width=0.75\linewidth]{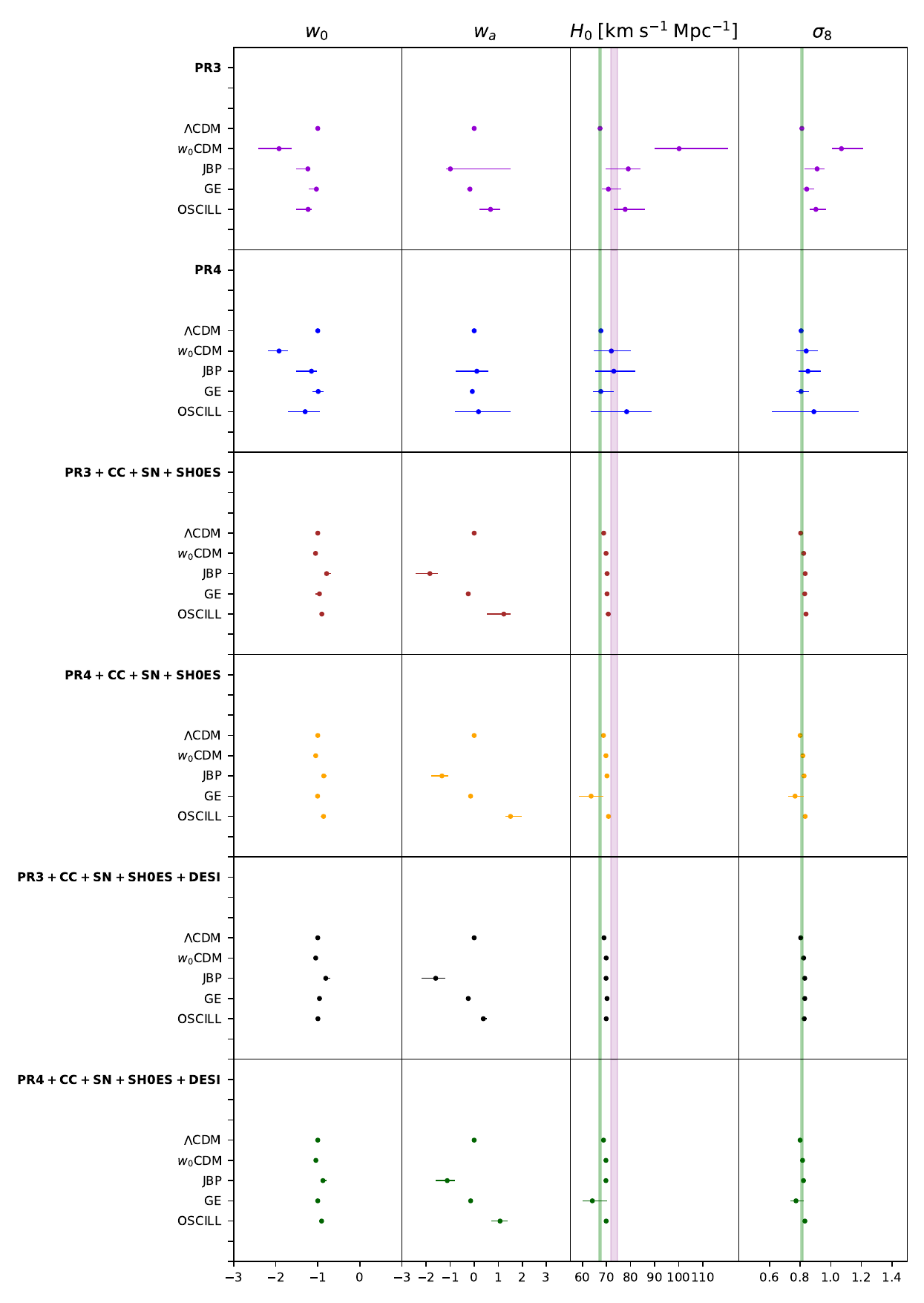}
    \caption{A whisker plot \cite{H0WhiskerPlot} depicting the results obtained from the four $w$CDM models and the $\Lambda$CDM model, when early-time data with combinations of late-time data that were used in section~\ref{sec:results}.}
    \label{fig:WHISK_ALL}
\end{figure}
A whisker plot, presented in Fig.~\ref{fig:WHISK_ALL}, illustrates the parameters $w_{0}$ and $w_a$, excluding the constant model, along with $H_0$ and $\sigma_8$. This visualisation considers early-time data and late-time with early-time data combinations, consistent with those applied in Section \ref{sec:results}. The vertical colored bars represent previous research values of $H_0$ and $\sigma_8$: the green bar indicates the Planck 2018 estimate of $H_0=67.4 \pm 0.5$ km s$^{-1}$Mpc$^{-1}$ and $\sigma_8=0.811 \pm 0.006$, while the pink bar corresponds to the SH0ES measurement of $ H_0= 73.2 \pm 1.3$ km s$^{-1}$Mpc$^{-1}$. Fig.~\ref{fig:WHISK_ALL} demonstrates that, overall, the constant parameterisation model exhibited larger uncertainties when relying solely on early-time data, compared to cases where combinations of late-time data were included. This highlights the necessity of late-time data for effectively constraining these four parameters. When late-time data was combined with \gls{CMB} data, the variation in the parameters $w_0$, $H_0$, and $\sigma_8$ across models diminished. Specifically, when Planck data was supplemented with late-time data, $w_0$ values clustered around $-1$, $H_0$ values centred around $70$, and $\sigma_8$ values ranged between $0.8$ and $0.9$. The logarithmic model (GE) demonstrated higher uncertainties and yielded lower values for $H_0$ and $\sigma_8$ compared to other models when PR4 data was used. While the quadratic parameterisation obtained high uncertainties for $w_a$, when only early-time data was considered. Among all models, the $w_0$CDM model recorded the highest uncertainties when solely utilising Planck 2018 data, particularly for the $H_0$ parameter. The Oscillatory model had obtained significantly large uncertainties with all four parameters when \gls{PR4} was used. As depicted in Fig.~\ref{fig:WHISK_ALL}, the $w_a$ parameter exhibited significant variation depending on the model and the data combination employed.

In conclusion, this section showed the $\Delta$\gls{AIC} and $\Delta$\gls{BIC} of the tested models, from them, it was deduced that with respect to the \gls{AIC} which model fit the data the best varied depending on the data combination while the \gls{BIC} showed that the Oscillatory parameterisation generally fit the data combinations better than the other chosen models. The \gls{AIC} and \gls{BIC} also showed that for certain data combinations, the tested models performed better than the $\Lambda$CDM model. Two overlapping graphs of the six $\Lambda$CDM parameters of the models and the standard model were generated, demonstrating that when early-time data was used, the chosen $w$CDM models do not affect the six $\Lambda$CDM parameters. However, when late-time data is added, deviations from the standard model are seen. It is important to note that despite this deviation, the tested models remain 1$\sigma$ away from the standard model. Two whisker plots were also shown, and it was concluded that the uncertainties decrease when early-time and late-time data are used together as opposed to using late-time or early-time data by themselves.

\section{Summary and Conclusions} \label{sec:conclusion}

Dark energy remains one of the greatest mysteries in modern cosmology, nearly two decades after its detection as the driving force behind the accelerated expansion of the Universe. The simplest and most widely accepted explanation is the $\Lambda$CDM cosmological model. While $\Lambda$CDM has proven successful in describing late-time cosmic acceleration and the Cosmic Microwave Background radiation, it has also raised significant tensions. The cosmological constant problem, related to discrepancies between theoretical predictions and observational estimates, remains one of the most profound challenges. Similarly, the cosmic coincidence problem, which questions why dark matter and dark energy densities are of comparable magnitude in the present cosmic epoch, has motivated the search for alternative models beyond $\Lambda$CDM. The pursuit of such alternatives gained momentum when discrepancies in $H_0$ estimations within $\Lambda$CDM were observed across different observational probes, exceeding 4 standard deviations. One potential approach to addressing these issues and because the precise nature of dark energy remains unknown, various forms of dark energy equation of state parameter that differs from $w = -1$ were explored. Specifically, four well-known dynamical dark energy models were investigated, aiming to assess their impact on the $H_0$ tension and the $\sigma_8$ tension in light of the latest Cosmic Microwave Background measurements from Planck 2020. In addition to the Planck data set, the Cosmic Chronometers, Pantheon+, Baryon Acoustic Oscillations, Dark Energy Spectroscopic Instrument, and local measurements of $H_0$ from \gls{SH0ES}, as well as earlier \gls{CMB} constraints from Planck 2018, to evaluate how cosmological constraints have evolved over time. \\
\begingroup
\begin{table}
\makebox[\textwidth][c]{
    \small
    \centering
    \begin{tabular}{lc}
    \hline
    \hline
         Name of $w$CDM parametrisation model& Equation of state of dark energy $w$\\
         \hline
         Constant model \dotfill & $w_{w0CDM}(a)=w_{0,w0CDM}$\\
         Quadratic model \dotfill & $w_{JBP}(a)= w_{0,JBP}+w_{a,JBP}~ a(1-a)$\\
         Logarithmic model \dotfill & $w_{GE}(a) = w_{0_{GE}} - w_{a_{GE}} \ln(a)$\\
         Oscillatory model \dotfill & $w_{OSCILL}(a)=w_{0,OSCILL} + w_{a,OSCILL} \left[ a \sin\left(\frac{1}{a}\right)-\sin(1)\right]$\\
         \hline
    \end{tabular}}
    \caption{A table showing the list of models that were considered and their respective equation of state of dark energy $w$.}
    \label{tab:equations_of_models}
\end{table}
\endgroup
Our analysis focuses on four dark energy models: the constant model; the Quadratic model; the Logarithmic model; and the Oscillatory model. The equation of state for each model can be found in Table~\ref{tab:equations_of_models}. Table~\ref{tab:w0CDM_valuesPR3PR4}-Table~\ref{tab:OSCILL_valuesPR3PR4+CCSN} present the best-fit and mean values of the six $\Lambda$CDM parameters, $w_0$, $w_a$, $H_0$, and $\sigma_8$. Additionally, the triangular plots for each table was plotted and displayed in this paper. Results indicate that for the constant model, the best-fit values of the dark energy equation of state parameter $w$, lie within the phantom regime. For the quadratic model, the logarithmic model, and the Oscillatory model, when combinations of early-time with late-time data were used, the mean values of $w$, lie within the quintessence regime at low redshifts but then shift to a phantom Universe at high redshifts. When only early-time data was used, the models showed a preference for a phantom Universe. When comparing the new datasets \gls{PR4} to the previously released data of \gls{PR3}, the models showed that generally, the \gls{PR4} data performed better than the older Planck data. However, when \gls{DESI} data was added to the data combinations, the models showed to constrain the parameters better than without it. There were some exceptions to this as the Oscillatory parameterisation, as when \gls{PR4} was considered by itself, the model obtained a completely degenerate $\sigma_8$, and the posteriors of $H_0$ as well as $w_{0,OSCILL}$ were larger than when \gls{PR3} was used. The oscillatory parameterisation also produced larger posteriors when \gls{PR4} was used together with late-time for the parameters $w_{0,OSCILL}$, and $w_{a,OSCILL}$ than when \gls{PR3} was used with late-time.\\

The Akaike Information Criterion (\gls{AIC}) and Bayesian Information Criterion (\gls{BIC}) statistics were used, and the values that were calculated for each model for each data combination were displayed in Table~\ref{tab:ALL_AIC&BIC}. From the values of the $\Delta$\gls{AIC}, it was deduced that when using Planck data with combinations of late-time data, the models fit the data better than the $\Lambda$CDM model. This indicates that the tested models could be statistically better than the standard model and could result in a better value of $H_0$ and $\sigma_8$ and lessen the tensions with the two parameters. When the values of the $\Delta$\gls{BIC} were analysed, it was concluded that generally the Oscillatory model had the lowest values out of the five parameterisation models, indicating that the Oscillatory model was statistically the best model out of all of them. Negative values of $\Delta$\gls{BIC} were found with the Logarithmic model, and the Oscillatory model when specific data combinations were taken.\\

Fig.~\ref{fig:Param_ALONE} and Fig.~\ref{fig:PARAM_ALL} were rendered to analyse the six $\Lambda$CDM parameters for all the chosen models with respect to the $\Lambda$CDM model when using \gls{PR4} only and then when using \gls{PR4}$+$\gls{CC}$+$\gls{SN$+$SH0ES}$+$\gls{DESI}, respectively. From the plots, it was decided that the chosen models did not have any effect on the six $\Lambda$CDM parameters when only early-time data was used. This implies that when using early-time data, the models do not affect the CMB power spectrum. When late-time data was added seen in Fig.~\ref{fig:PARAM_ALL}, the six $\Lambda$CDM parameters obtained from the tested models are no longer the same as the standard model. However, the deviation from the standard model is 1$\sigma$ for all the $w$CDM models showing that even though there is a deviation from the standard model, all the tested models limit to the $\Lambda$CDM model for the six $\Lambda$CDM parameters.\\

A whisker plot was generated, as seen in Fig.~\ref{fig:WHISK_ALL}, showing the $w_0$, $w_a$, $H_0$ and $\sigma_8$ in the last figure for all data combinations that were used in section~\ref{sec:results}. From the figure, it was concluded that the chosen models obtained large uncertainties when Planck data themselves are used. However, when late-time and early-time data are used together, the uncertainties are significantly lessened. This showed that late-time data is required for the $w$CDM models to give accurate values of the parameters and consequently an accurate description of the early and late Universe.\\

In summary, the analysis of this study showed how the $w$CDM models do not deviate so much from the $\Lambda$CDM model. It was seen that the tested models fit the older Planck data better than the newer one. This could be due to the newer model having less uncertainties than its predecessor. This study indicates that the tested dark energy models can align with the observational data. However, Bayesian analysis suggests that $\Lambda$CDM model remains the preferred model. A potential direction for further research could involve exploring a more general framework in which the sound speed of dark energy varies instead of remaining constant. This approach may shed light on the fundamental characteristics of dark energy models, and is reserved for future investigation. The code, figures, and values computed in this research can be found in the Github link: https://github.com/kathleenUni/Update-of-wCDM-models/blob/main/README.md .

\begin{acknowledgments}
This article is also based upon work from COST Action CA21136 Addressing observational tensions in cosmology with systematics and fundamental physics (CosmoVerse) supported by COST (European Cooperation in Science and Technology). It was funded by the ENDEAVOUR II scholarship scheme offered by the Government of Malta Ministry for Education, Sport, Youth, Research and Innovation (MEYR). MEYR was not involved in any capacity in relation to the study design, collection, analysis and interpretation of data, writing of the report and decision to submit the article for publication. KS would also like to acknowledge Glenn Wallace for the help in the implementation of the Oscillatory parameterisation.
\end{acknowledgments}

\newpage

\listoffigures
\listoftables
\newpage

\section*{Glossary}
\printglossary[title=]

\newglossaryentry{BAO}{
    name=BAO,
    description={Baryonic
Acoustic Oscillations}
}

\newglossaryentry{CC}{
    name = CC,
    description = {Cosmic Chronometers}}

\newglossaryentry{CMB}{
    name = CMB,
    description = {Cosmic Microwave Background}
}

\newglossaryentry{PR4}{
    name = PR4,
    description = {Planck 2020}
}

\newglossaryentry{PR3}{
    name = PR3,
    description = {Planck 2018}
}

\newglossaryentry{SH0ES}{
    name = SH0ES,
    description = { Supernovae H0 for the Equation of State}
}

\newglossaryentry{SN}{
    name = SN,
    description = { Type 1a Supernovae}
}

\newglossaryentry{FLRW}{
    name = FLRW,
    description = {Friedmann-Lema$\hat{i}$tre-Robertson-Walker}
}

\newglossaryentry{MCMC}{
    name = MCMC,
    description = {Markov Chain Monte Carlo}
}

\newglossaryentry{CLASS}{
    name = CLASS,
    description = {Cosmic Linear Anisotropy Solving System}
}

\newglossaryentry{AIC}{
    name = AIC,
    description = {Akaike Information Criterion}
}

\newglossaryentry{BIC}{
    name = BIC,
    description = {Bayesian Information Criterion}
}

\newglossaryentry{DES}{
    name = DES,
    description = Dark Energy Survey
}

\newglossaryentry{SDSS}{
     name = SDSS,
     description = {Sloan Digital Sky Survey}
}

\newglossaryentry{LRG}{
    name = LRG,
    description = {Luminous Red Galaxies}
}

\newglossaryentry{ELG}{
    name = ELG,
    description = {Emission-line Galaxies}
}

\newglossaryentry{DESI}{
    name = DESI,
    description = {Dark Energy Spectroscopic Instrument}
}

\newglossaryentry{JBP}{
    name = JBP,
    description = {Jassal-Bagla-Padmanabhan}
}

\newglossaryentry{GE}{
    name = GE,
    description = { G.
Efstathiou}
}

\newglossaryentry{SN$+$}{
    name = SN$+$,
    description = {Pantheon+}
}

\newglossaryentry{SN$+$SH0ES}{
    name = SN$+$SH0ES,
    description = {Pantheon$+$ and  Supernovae H0 for the Equation of State}
}

\newglossaryentry{BOSS}{
    name = BOSS,
    description = {Baryon Oscillation Spectroscopic Survey}
}
\newglossaryentry{PR3$/$ALL}{
    name = PR3$/$ALL,
    description = {PR3$+$CC$+$SN+SH0ES$+$DESI}
}
\newglossaryentry{PR4$/$ALL}{
    name = PR4$/$ALL,
    description = {PR3$+$CC$+$SN+SH0ES$+$DESI}
}

\newglossaryentry{ALL}{ 
    name = ALL,
    description = {CC$+$SN+SH0ES$+$DESI}}

\newpage

\bibliographystyle{apalike}
\setkeys{acs}{articletitle = true}
\bibliography{references}

\end{document}